\documentclass[12pt,a4paper]{article}

\usepackage{amsmath,amssymb,amsfonts}
\usepackage{graphicx,epsfig}
\usepackage{color}
\usepackage{amsmath}
\usepackage{amsfonts}
\usepackage{amssymb}
\usepackage{float}
\usepackage{cancel}
\usepackage{caption}
\usepackage{graphics}
\usepackage{hyperref}
\usepackage{graphicx}
\usepackage{feynmf}
\usepackage{makeidx}
\usepackage{subfigure}
\usepackage{dsfont}
\usepackage{bbold}
\usepackage{bbm}
\usepackage{slashed}

\ifx\pdfoutput\undefined
\usepackage[dvips,bookmarks]{hyperref}	
\else
\usepackage{hyperref}	
\fi
\hypersetup{colorlinks,bookmarksopen,bookmarksnumbered,citecolor=rossos,
linkcolor=rossos,pdfstartview=FitH,urlcolor=rossos}

\def\circa#1{\,\raise.3ex\hbox{$#1$\kern-.75em\lower1ex\hbox{$\sim$}}\,}

\numberwithin{equation}{section} 
\setcaptionmargin{1cm}

\usepackage[table]{xcolor}

\definecolor{grigino}{cmyk}{0,0,0,0.2}
\definecolor{mentuccia}{cmyk}{0.4,0,0.3,0.1}
\definecolor{arancino}{cmyk}{0,0.1,0.4,0}
\definecolor{menta}{cmyk}{0.7,0,0.5,0.3}
\definecolor{grigios}{cmyk}{0,0,0,0.5}
\definecolor{bianco}{cmyk}{0,0,0,0}
\definecolor{arancio}{cmyk}{0,0.2,0.6,0}
\definecolor{grigio}{cmyk}{0,0,0,0.1}
\definecolor{rosa}{cmyk}{0,0.1,0.1,0.02}
\definecolor{rosino}{cmyk}{0,0.05,0.05,0.02}
\definecolor{rosas}{cmyk}{0,0.3,0.25,0.05}
\definecolor{celeste}{cmyk}{0.1,0,0,0.02}
\definecolor{giallino}{cmyk}{0,0,0.4,0.02}
\definecolor{rosso}{cmyk}{0,1,1,0.4}
\definecolor{rossos}{cmyk}{0,1,1,0.55}
\definecolor{rossoc}{cmyk}{0,1,1,0.2}
\definecolor{blu}{cmyk}{1,1,0,0.3}
\definecolor{blus}{cmyk}{1,1,0,0.5}
\definecolor{bluc}{cmyk}{1,1,0,0.1}
\definecolor{blucc}{cmyk}{0.7,0.5,0,0}
\definecolor{viola0}{cmyk}{0,0.4,0,0.04}
\definecolor{viola}{cmyk}{0,0.5,0,0.05}
\definecolor{viola2}{cmyk}{0,1,0.2,0.6}
\definecolor{verde}{cmyk}{0.92,0,0.59,0.25}
\definecolor{verdec}{cmyk}{0.92,0,0.59,0.15}
\definecolor{verdecc}{cmyk}{0.42,0,0.8,0.05}
\definecolor{verdes}{cmyk}{0.92,0,0.59,0.4}
\definecolor{verdino}{cmyk}{0.12,0,0.3,0.02}
\definecolor{giallo}{cmyk}{0,0,1,0}
\definecolor{gialloverde}{cmyk}{0.44,0,0.74,0}

\newcommand{\HH}{\mathcal{H}}
\newcommand{\GG}{\mathcal{G}}

\newcommand{\be}{\begin{equation}}
\newcommand{\ee}{\end{equation}}
\newcommand{\ben}{\begin{equation*}}
\newcommand{\een}{\end{equation*}}
\newcommand{\ba}{\begin{eqnarray}}
\newcommand{\ea}{\end{eqnarray}}
\newcommand{\tr}{\mathrm{tr}}

\newcommand{\GeV}{~\mathrm{GeV}}
\newcommand{\TeV}{~\mathrm{TeV}}
\newcommand{\gev}{\textrm{GeV}}
\newcommand{\tev}{\textrm{TeV}}

\begin{document}

\begin{titlepage}
\begin{flushright}
\end{flushright}
\vskip 1.0cm
\begin{center}
{\huge\color{rossos} \bf T-parity, its problems and their solution} \vskip 1.0cm
 {\large 
Duccio Pappadopulo$^{a}$
 \ \ and\ \  Alessandro Vichi$^a$
} \\[1cm]
{\it $^a$ Institut de Th\'eorie des Ph\'enom\`enes Physiques, EPFL,  CH--1015 Lausanne, Switzerland}\\[5mm]
\vskip 1.0cm 

\abstract{We point out a basic difficulty in the construction of little-Higgs models with $T$-parity which is overlooked by large part of the present literature. Almost all models proposed so far fail to achieve their goal: they either suffer from sizable electroweak corrections or from a breakdown of collective breaking. We provide a model building recipe to bypass the above problem and apply it to build the simplest $T$-invariant extension of the Littlest Higgs. Our model predicts additional $T$-odd pseudo-Goldstone bosons with weak scale masses.}

\end{center}
\end{titlepage}

\section{Introduction}

The Large Hadron Collider will soon shed light on the dynamics of electroweak symmetry breaking (EWSB). The elementary Higgs sector that defines the Standard Model (SM) is unsatisfactory as it is affected by the hierarchy problem. On the other hand generic extensions of the SM dealing with this issue have to face some tension between the scale where new physics appears to restore naturalness and the bounds on this scale indicated by flavour physics and electroweak precision tests (EWPT). This tension is referred as the little hierarchy problem.

Little-Higgs (LH) models \cite{minmoose,littlest, antisym, so5moose, su9su8, so9} address the little hierarchy problem generating the Fermi scale without fine tuning from a strong coupling scale $\Lambda=4\pi f\sim 10~\tev$ through the mechanism of collective breaking. In this setup the Higgs boson is the pseudo-Goldstone boson of some global symmetry broken at the scale $f$. In this class of models the solution of the big hierarchy problem is postponed to a UV-completion at the 10~\tev scale.

In trying to apply this idea to specific constructions one realizes that none of the LH models with a single light scalar doublet is fully compatible with the EWPT without relying on some fine tuning of parameters.
The simplest LH models indeed suffer from several problems due to tree level contributions coming from heavy states \cite{lhproblems}.  The corrections induced, for instance, in the electroweak oblique parameters are parametrically given by \cite{lep2aftermath, lhstrumia}
\be\label{estimatelh}
\delta \hat S,~\delta \hat T\sim \left(\frac{m_W}{m'_W}\right)^2\sim\frac{\alpha_{SM}}{4\pi}
\ee
where $m'_W\sim g_{SM}f$ is the mass of the new heavy vectors. Numerical coefficients found in definite models can make these corrections acceptable. Even so, the naive estimate shows they are expected to be on the verge of the experimental limit.

These remnants of the little hierarchy problem can be observed in specific models. In \cite{minmoose}, for instance, there is no custodial symmetry for the Higgs doublet and custodial breaking operators come directly from the $\sigma$-model. The coset in \cite{littlest} is custodial symmetric but the gauging of the partner of the hypercharge again generates large corrections to the $\hat T$ parameter unless the scale $f$ is unnaturally large. Further problems come from unavoidable VEVs of triplets scalars sourced by tadpoles after EWSB. More complex models with larger cosets like \cite{su9su8, so9} are also at stake due to the presence of destabilizing singlets as pointed out in \cite{dangsinglets}\footnote{Contrasting with these various difficulties it is worth mentioning \cite{bestest} as a recent new twist in LH model building.}.

The introduction of an additional discrete symmetry, $T$-parity, is a further attempt to control the size of the corrections coming from new physics. The goal of little-Higgs models with $T$-parity (LHT) \cite{tparity1, tparity2, tparity3, perel} is to obtain a spectrum where only the SM states are $T$-even. All the tree-level processes mediated by heavy ($T$-odd) particles are thus automatically forbidden, together with the potential tadpoles sourcing triplet VEVs. This changes the estimate in \ref{estimatelh} to
\be\label{estimatelht}
\delta \hat S,~\delta \hat T\sim\left(\frac{\alpha_{SM}}{4\pi}\right)^2,
\ee
thus relieving any major tension with the EWPT.

The implementation of $T$-parity is clean in the $\sigma$-model and gauge sector. It relies on the observation that a symmetric coset $\mathcal{G}/\mathcal{H}$, 
\be
\ [T,\,T]\sim T\,, \quad [T,\, X]\sim X\,, \quad [X,\, X]\sim T\qquad T\in \mathcal H,\,\, X\in \mathcal G/\mathcal H
\ee
posses an intrinsic $Z_2$-parity
\be\label{parity}
\ T\longrightarrow +T \,,\qquad X\longrightarrow -X\,,
\ee
which preserves the $\mathcal{G}$ algebra\footnote{With this definition $T$-parity makes all the Goldstones, the Higgs doublet in particular, $T$-odd and is spontaneously broken after EWSB. In explicit models an unbroken combination of \ref{parity} and electroweak finite rotations is used so as to realize an even Higgs field.}.

The general setup we consider is a little-Higgs theory based on a symmetric coset $\mathcal{G}/\mathcal{H}$. We gauge a subgroup $\mathcal{G}_1\times\mathcal{G}_2$ which is broken to the vector subgroup $\mathcal{G}_{V}\subset \mathcal H$ at the scale $f$. The discrete symmetry exchanges $\mathcal{G}_1$ and $\mathcal{G}_2$ generators resulting in $T$-even unbroken $T$-odd broken generators.

The SM fermions have to correspond to even eigenvectors of $T$. The above discussion suggests that it is not possible to introduce fermions charged only under, say, $\mathcal{G}_1$ since their coupling would violate $T$-parity. The simplest thing to try is to embed the SM fermions into representations of the unbroken group $\HH$. We will refer to these kind of representations as ``composite'', since these fields are strongly coupled to the Goldstone bosons at the scale $\Lambda$. On the other hand we define as ``elementary'' those fields belonging to representations of the broken symmetry group $\mathcal G$.

The simple choice of using composite SM fermions introduces undesired effects in the low energy theory. The kinetic term of a composite fermion is dictated by the CCWZ construction \cite{CCWZ} and contains interactions involving the Goldstones (among which the Higgs doublet) and the SM fermions. As we will show in the next section these generate order $(v/f)^2$ corrections to neutral and charged currents at tree level and quadratically divergent  four-fermion operators at one-loop order \cite{tparity2}. The vertex corrections, in particular, have been overlooked in the existing literature. Both these effects are subject to stringent constraints from experiments, pointing to a large value of $f$ in conflict with naturalness.
 
To avoid these effects the SM fermions can be taken as a $T$-even combination of elementary fermions sitting in some, if necessary incomplete, representation of the group $\GG$ \cite{tparity3}. 
One automatically gets rid of the vertex corrections and of the divergent four-fermions interactions since the $\GG$ symmetry is linearly realized.

As we will explain in the following sections this option has the drawback of introducing a non-minimal fermion content in the theory. We will point out that giving a mass to the extra states 
leads to a tension between $T$-parity, $SU(2)$ gauge invariance and the collective breaking mechanism\footnote{This issue was realized already in \cite{csaki}, where a solution is proposed which is, however, not easy to disentangle from the ambitious attempt to build an UV complete model.}.
Moreover the new heavy states can in general mix with the SM fermions and give rise to intolerable vertex corrections. These effects have a different origin from those mentioned in the context of composite fermions. 
Specializing to the littlest Higgs model \cite{littlest} we find that the coset has to be enlarged to overcome these difficulties. The model we propose is characterized by the presence of a singlet and a triplet of $T$-odd pseudo-Goldstones with small masses in the 100~\gev ~range. The singlet is thus a natural dark matter candidate.
 
The work is organized as follow: in section \ref{tparityproblems} we clarify the difficulties in embedding the SM fermions into an LHT model. 
We explore some proposals present in the literature explaining their problems. In section \ref{mincoset} we propose an LHT model addressing the issues we presented in section \ref{tparityproblems} and we briefly discuss its phenomenology, finally we conclude in section \ref{conclusion}.

\section{Fermions: composite or elementary?}\label{tparityproblems}

In what follows we consider models in which a simple group $\GG$ is spontaneously broken to a subgroup $\HH$ in such a way that $\GG/\HH$ is a symmetric space\footnote{Another possibility discussed in the literature is the ``moose'' structure where the global group is the product of several simple factors $\GG$, broken down to the diagonal combination. Our considerations will apply also to this case.}. We gauge a subgroup of $\GG_1\times \GG_2\subset \GG$ which is broken to the diagonal group $\GG_V\subset \HH$. We call $T^a$ the generators of $\HH$ and $X^a$ those spanning the coset $\GG/\HH$. 

After the breaking the SM and heavy vectors are identified with the following linear combinations
\be
\	A_{SM}^a=\frac{A_{1}^a+A_{2}^a}{\sqrt2},\qquad A_{H}^a=\frac{A_{1}^a-A_{2}^a}{\sqrt2}.
\ee 
These are respectively even and odd under the $Z_2$ symmetry exchanging the generators of $\mathcal G_1$ with those of $\mathcal G_2$. Such $Z_2$ parity is the restriction to $\mathcal G_1\times \mathcal G_2$ of the automorphism $\mathcal A$ defined on the symmetric space $\mathcal G/\mathcal H$
\be\label{eq:automorphism}
\mathcal A: T^a\rightarrow T^a, \qquad X^a \rightarrow -X^a.
\ee

In order to preserve the above symmetry even in presence of the gauging the equality of the two gauge couplings $g_1=g_2$ must be enforced. Moreover, under $\mathcal A$ the Goldstone boson matrix transforms as
\be
\mathcal A: \Pi=\Pi^a X^a\rightarrow -\Pi\,.
\ee
As mentioned in the introduction, the $T$-parity defined in this way makes all the Goldstone bosons odd and is spontanously broken after EWSB. This is fixed defining the action of $T$-parity as the composition of the automorphism $\mathcal A$ with an $SU(2)_W$ rotation  
\be
T:\quad \Pi\rightarrow -\Omega \Pi \Omega\,,\qquad \Omega=-\exp 2\pi i Q_W^3\,,
\ee
where $Q_W^3$ is the third generator of $SU(2)_W$.
%
%
Since $\Omega$ corresponds to an $SU(2)$ rotation of $2\pi$ around the third axis the total effect of $T$-parity is to change sign to the $SU(2)$ representations with integer isospin. Thus, if the coset $\GG/\HH$ decomposes as a set of triplets, doublets and singlets only the doublets will be even under $T$-parity. 

Let us now consider a set of composite fermions $\Psi$ sitting in a representation of the subgroup $\HH$. By definition these fields transform non-linearly under $\GG$:
\be
\Psi\rightarrow h\Psi,\qquad h\in\HH,
\ee
\be
\Psi\rightarrow h(g,\xi)\Psi,\qquad g\in\GG,
\ee
where $h(g,\xi)$ is the non linear transformation associated to $g$ containing a dependence on the Goldstone bosons matrix $\xi=e^{i\Pi}$. The relation defining $h$ is (\cite{CCWZ})
\be
\GG: \,\xi\rightarrow \xi'=g\,\xi\, h^\dagger(g,\xi)= h(g,\xi)\, \xi \,\tilde g^\dag\,,
\ee
where, schematically, $g=e^{i(T+X)}$ while $\tilde g=e^{i(T-X)}$.\\
Even in the absence of gauging the kinetic term of a composite field is non-trivial since the non-linear transformation $h(g,\xi)$ contains an explicit dependence on the coset coordinates. We define a covariant derivative making use of the Maurer-Cartan vector $\mathcal E_\mu\equiv  \mathcal E_\mu^a T^a$ \cite{CCWZ}:
\be\label{eq:Maurer-Cartan}
\xi^\dagger\partial_\mu \xi =i \mathcal E_\mu^a T^a + i\mathcal D_\mu^a X^a\nonumber, \qquad i\mathcal E_\mu = \tfrac12\left(\xi^\dagger\partial_\mu \xi+\xi \partial_\mu \xi^\dagger \right)
\ee
The covariant object is  $\partial_\mu \Psi+i\mathcal E_\mu \Psi$ which transforms as $\Psi$ under $\GG$.
At this point we can deduce the transformation properties of $\Psi$ under $T$-parity. Since
\be
T:\, \mathcal E_\mu \rightarrow \Omega \mathcal E_\mu \Omega
\ee
the request that $i \bar \Psi (\displaystyle{\not\,}\partial+i\displaystyle{\not}\mathcal E)\Psi$ is invariant forces the choice
\be\label{eq:compositeTtransf}
T:\,\Psi\rightarrow \pm \Omega \Psi
\ee
The above relation has an important consequence for the phenomenology of LHT: as we discuss in section \ref{troublescomposite} a model with composite fermions may contain even states that mix with the SM fermions, introducing large vertex corrections (see fig.\ref{fig:vertexcorrectionB}). Notice that the transformation \ref{eq:compositeTtransf} is again, up to a sign, an $SU(2)_W$ rotation; hence the lagrangian is not required to be invariant under additional symmetries: up to now only the invariance under $\mathcal A$ has been imposed.

Let us now discuss the inclusion of representations of the global symmetry group $\mathcal G$. $T$-parity invariance requires the use of irreducible representations of the semi-direct product group $\mathcal G\rtimes  Z_2$, where $Z_2$ acts on $\mathcal G$ as the $T$-parity automorphism $\mathcal A$. Now, by assumption, $T$-parity coincides with the automorphism defining the symmetric space over which a specific LHT model is built. The classification of these spaces and the action of $T$-parity on the irreducible representations of the global group can be found in \cite{slansky}.

In the particular case of $SU(5)/SO(5)$, the coset used in the littlest Higgs model, the action of $\mathcal{A}$ is just the complex conjugation. This can be checked explicitly by a change of basis in the Lie algebra
\be
T'_A=U T_A U^{-1},\qquad 
U=\frac1{\sqrt2}\begin{pmatrix}
 1_2 & & 1_2\\
&\sqrt 2 &\\
i 1_2 & & -i 1_2
\end{pmatrix}.
\ee
In this new basis the broken generators are real while the unbroken ones are purely imaginary. The semi-direct group $SU(5)\rtimes Z_2$ is thus represented in this case using $\mathbf r\oplus \overline{\mathbf r}$ representations if $\mathbf r$ is complex, or just $\mathbf r$ if it is real.

One further detail must be settled. In order to define $T$-parity for the Goldstone bosons we composed the automorphism $\mathcal A$ with a finite $SU(2)_W$ rotation $\Omega$. This has to be done consistently also for the fields $\Phi$ belonging to representations of $\mathcal G$
\be
T:\qquad \Phi\rightarrow\pm \Omega \mathcal A(\Phi),
\ee
where with an abuse of notation we called in the same way the finite $SU(2)$ rotation and the matrix representing it. Notice that replacing $\Omega$ with a matrix $P$ such that $P^2=1$ is in general asking more than the basic requirement of $\mathcal A$-invariance and $SU(2)_W$ invariance, in particular the transformation
\be\label{pparity}
T':\qquad \Phi\rightarrow P \mathcal A(\Phi),
\ee
may not be a symmetry.

%
%

Finally, the doubling of the field content can be sometimes avoided for scalars. For instance, in the littlest Higgs model, the matrix of Goldstone bosons is transformed into its hermitian conjugate up to a change of basis. 
For fermions this is usually not possible: if we want the Poincar\'e group to commute with the internal symmetries, two representations connected by a $T$-parity transformation must have the same Lorentz quantum numbers.

\subsection{Troubles with composite fermions}\label{troublescomposite}

In early models of LHT the SM fermions are chosen to be part of a representation of the low energy group $\HH$ \cite{tparity1, tparity2}. 
This choice appears to be the most natural one given the fact that representations of the unbroken subgroup can be eigenstates of T-parity. 
We want to illustrate the reasons why this kind of construction is not viable. One of these is well known, while another, to our knowledge, has been overlooked.

\begin{figure}[t] 
\centering%
\subfigure[\label{fig:vertexcorrectionA}]%
{\includegraphics[scale=0.7]{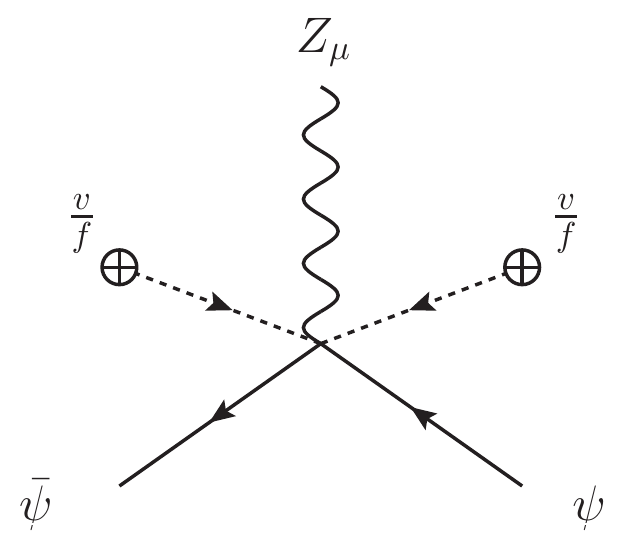}}\qquad\qquad\qquad 
\subfigure[\label{fig:vertexcorrectionB}]%
{\includegraphics[scale=0.7]{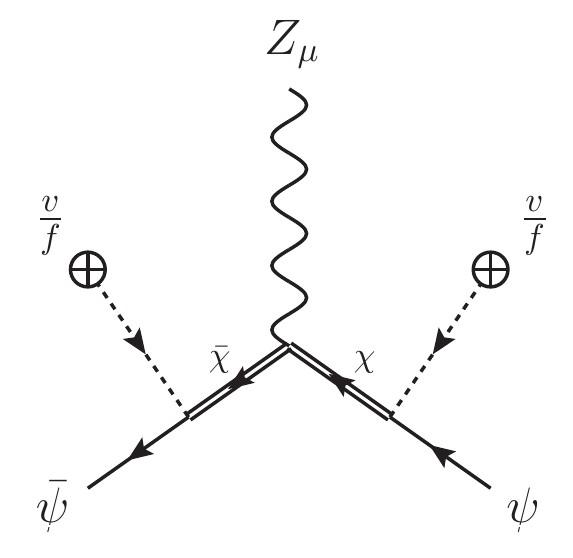}}\qquad\qquad
\caption{\small Vertex corrections from composite fermions\label{fig:vertexcorrection}} 
\end{figure}

As discussed in the previous section a composite representation transforms non-linearly under the global group $\GG$ and the CCWZ construction must be used to write an invariant kinetic term.
When the subgroup $\GG_1\times \GG_2 \subset \GG$ with generators $Q^a_1$ and $Q_2^a$ is gauged, the covariant derivative becomes
\be\label{eq:covariant derivative with connection}
D_\mu\Psi = \partial_\mu \Psi +i\mathcal E_\mu\Psi -\frac{ig}{2}\sum_{j=1,2}\left( \xi^\dagger  A_{j\,\mu}^a Q^a_j\xi +\xi  A_{j\,\mu}^a \widetilde Q^a_j\xi^\dag \right) \Psi
\ee
where schematically $Q= T+X$ and $\widetilde Q= T-X$.
The axial combination of the gauge fields acquires a mass from the Goldstone boson kinetic term, hence the light vectors $A_{1\mu}^a+A_{2\mu}^a$ can be identified with the usual SM gauge vectors (in all the models the residual $\GG_V$ contains the electroweak gauge group). The CCWZ kinetic term for the composite fermion contains interactions among gauge vectors, Goldstone bosons and the fermion. After EWSB these couplings introduce potentially dangerous vertex corrections in the way depicted in fig.\ref{fig:vertexcorrectionA}.

%
%
These can be calculated in explicit models, e.g. \cite{tparity2}, and are generically expected to modify the $Zff$ vertex by an amount $\delta g/ g\sim v^2/f^2$ leading to unacceptable modifications to several $Z$-pole observables unless $f$ is tuned to high values.
This feature is shared by all the models employing composite SM fermions. \newline

\begin{figure}[t] 
\centering%
\subfigure[\label{fig:flavor}]%
{\includegraphics[scale=0.6]{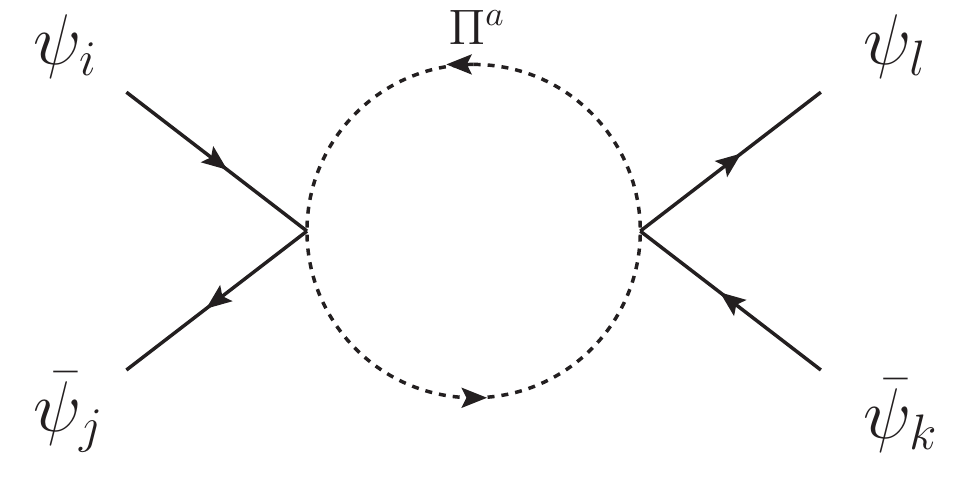}}\qquad \qquad 
\subfigure[\label{fig:incomplete}]%
{\includegraphics[scale=0.7]{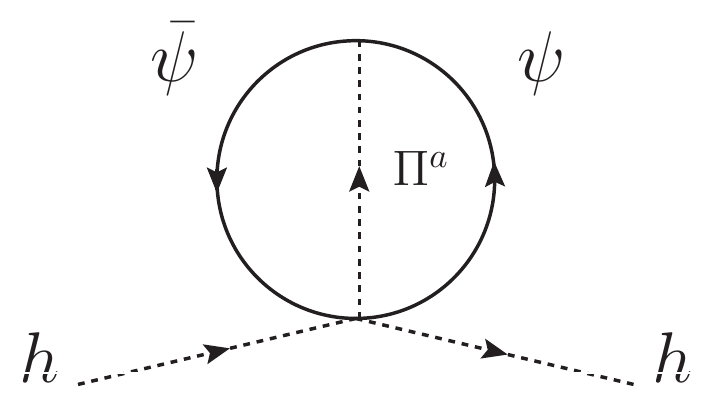}}\qquad\qquad 
\caption{\small Quadratically divergent diagram contributing to four-fermion operators (a) and quartic divergent contribution to the Higgs boson mass (b) generated by the CCWZ kinetic term of a composite fermion.} 
\end{figure}

The vertex corrections discussed so far originate from the coupling of the fermions with the Higgs field via their $\mathcal G$-invariant kinetic terms. The CCWZ term introduces at the same time a derivative interaction with all the other heavy Goldstone bosons such as $\bar\psi\gamma^\mu\psi \pi D_\mu\pi$, $\bar\psi\gamma^\mu\psi H^\dagger D_\mu H$. These generate quadratically divergent four-fermion operators through the diagram in fig.\ref{fig:flavor} \cite{tparity3}, the size of which is estimated to be, according to naive dimensional analysis (NDA):
\be\label{O4F}
\mathcal O_{J_1J_2}\approx  \Lambda^2 f^2 \frac{J^\mu_1}{f^2\Lambda} \frac{J^\mu_2}{f^2\Lambda}, \qquad J_I^\mu=\bar\psi_I \gamma^\mu\psi_I.
\ee
Strong bounds on these compositness effects come from LEP searches. The more constrained among the contact operators is the $eedd$ one; following \cite{PDG4f}, the scale $M$ appearing in
\be
\pm\frac{1}{M^2}\overline e_L\gamma_\mu e_L\overline d_L\gamma^\mu d_L
\ee
has to satisfy $M\gtrsim 10.5\TeV$. The exact form of the lower bound on $f$ depends on the particular model\footnote{In \cite{perel} the calculation of the size of these four-fermion operators has been carried out for a specific realization of the littlest Higgs with $T$-parity. There the SM fermions are elementary and the quadratic divergence in \ref{O4F} is cutted off by the presence of extra heavy fermions. If these heavy fermions are pushed to the $\sigma$-model cutoff then the SM fermions become composite and $f$ is bounded by $f\gtrsim 2.5\TeV$, so that a small numerical factor can generically ameliorate the naive estimate.}, nevertheless the constraint coming from  the appearance of interactions like \ref{O4F} always introduces a large fine-tuning in the theory.

Another issue related to the presence of composite fermions was realized in \cite{tparity2}. It regards the presence in the theory of incomplete representations of $\mathcal H$.
An incomplete representation, either of $\GG$ or of $\HH$, breaks explicitly the global group $\GG$. In the former case the breaking of the symmetry is transmitted to the Goldstone sector by interactions proportional to some parameter $\lambda$ (a gauge or a Yukawa coupling for instance). As long as $\lambda$ is perturbative we can keep under control the explicit breaking. On the other hand for an incomplete representation of $\mathcal H$ there is no small parameter since non invariant interactions come directly from the fermion kinetic term. This inevitably results in contributions to the Higgs boson mass\footnote{There may be exceptions to this if the unbroken group is so big to allow for incomplete representations that leave enough of the global symmetry unbroken. Since the breaking of the global symmetry due to incomplete composite representations is not controlled by any small parameter, any justification of collective breaking in models employing such representations must be substantiated by symmetry considerations. This statement apply for instance to the model in  \cite{wyler}. There incomplete representations are used but no quadratic divergences appear up to two loops. Though we have not explicitly checked we expect this either to be the effect of some symmetry being at work or just an accident. In this latter case we expect divergences to ultimately appear at higher orders.}.

In order to see this explicitly let us consider again the littlest Higgs with $T$-parity and an incomplete $SO(5)$ representation $\Psi=(\psi,0, 0)$. Its kinetic term is defined with the covariant derivative in eq.\ref{eq:covariant derivative with connection}.  Since $\Psi$ is incomplete this term preserves just the gauge symmetry. Indeed, as already pointed out in \cite{tparity2}, there are two-loop divergences contributing with terms $\mathcal O(\Lambda^2)$ to the Higgs boson mass through diagrams such as those showed in Fig \ref{fig:incomplete}.  

Similar problems have to be faced in trying to decouple a component of a composite representation, e.g. a singlet. 
Working again in the littlest Higgs framework, we take a complete $SO(5)$ multiplet $\Psi=(\psi_1,\chi,\psi_2)$ and we consider the following schematic model:
%
%
\be\label{eq:nonSO5inv}
\mathcal{L}=i\bar{\Psi} \displaystyle{\not} D  \Psi+  i \bar{\chi'} \displaystyle{\not} D  \chi' +M_S \bar{\chi}\chi'.
\ee
We take $\chi'$ to be an $SU(2)_W$ singlet with the same $T$-parity of $\chi$. Its covariant derivative contains only the hypercharge field. The last term in (\ref{eq:nonSO5inv}) breaks explicitly the $SO(5)$ invariance and we expect contributions to the Goldstone bosons masses. In the limit of vanishing $M$ the symmetry is restored and all the new divergences must be proportional to this parameter.  Indeed, the kinetic term for $\Psi$ contains a derivative coupling between two fermions and two Higgs scalars that generates, through the diagram in fig. \ref{fig:incomplete} an $\mathcal O(M^2)$ contribution
%
to the Higgs boson mass. 
 
\subsection{Elementary fermions and the SM spectrum}

The effects described in Sec.2.1 constrain the appearance of composite representations in a LHT model and exclude the possibility of using composite representations to embed the SM fermions. 

In \cite{tparity3, perel} the SM fermions are therefore regarded as a $T$-even combination of elementary fermions sitting in some, if necessary incomplete, representation of the global group $\GG$.

In order to discuss the main obstacles one has to overcome in the construction of a successful model we will work in the specific framework given by the littlest Higgs. The considerations we make will though be valid also in other realizations.

The symmetry breaking pattern we consider \cite{littlest} is described by a $5\times5$ matrix $\Sigma$ transforming in the symmetric representation of $SU(5)$: $\Sigma\rightarrow g\Sigma g^T$. $\Sigma$ can be written in terms of the Goldstone matrix $\Pi$ and the vev $\Sigma_0$ as 
\be
\Sigma=e^{i\frac{\Pi}{f}}\Sigma_0 e^{i\frac{\Pi^{T}}{f}}=e^{2i\frac{\Pi}{f}}\Sigma_0\equiv \xi^2 \Sigma_0\,,
\ee
with 
\be
\Pi=\Pi^a X^a\,,\qquad 
\Sigma_0=\begin{pmatrix} & & \mathbb 1_2\\
&1&\\
\mathbb{1}_2& &\end{pmatrix}\,,
\ee
where we used the relations
\ba\label{gen}
T^a\Sigma_0+\Sigma_0T^{a T}&=&0, \\
X^a\Sigma_0-\Sigma_0X^{a T}&=&0.
\ea
The explicit expression for $\Pi$ in terms of the Goldstone bosons is provided in Sec.3, where an explicit model is presented. 

We define the $T$-parity as the composition of the automorphism $\mathcal A$ and an $SU(2)_W$ rotation $\Omega=\mbox{diag}(1,1,-1,1,1)$:
\be
T: \quad\Pi\rightarrow - \Omega \Pi \Omega.
\ee
To implement collective breaking in the gauge sector  two $SU(2)\times U(1)$ copies are gauged inside $SU(5)$:
\be
Q_1^a=\begin{pmatrix}
\sigma^a/2 & 0 & 0\\
0 & 0 &0 \\
0 & 0 &0 
\end{pmatrix},\qquad Y_1=\textrm{diag}\left(\tfrac{3}{10},\tfrac{3}{10},-\tfrac{2}{10},-\tfrac{2}{10},-\tfrac{2}{10}\right),
\ee
\ben
Q_2^a=\begin{pmatrix}
0 & 0 & 0\\
0 & 0 &0 \\
0 & 0 & -\sigma^{a*}/2 
\end{pmatrix},\qquad Y_2=\textrm{diag}\left(\tfrac{2}{10},\tfrac{2}{10},\tfrac{2}{10},-\tfrac{3}{10},-\tfrac{3}{10}\right).
\een

For each SM doublet we introduce a couple of doublets which we embed in incomplete representations\footnote{$SU(5)$ invariance is explicitly broken by the appearance of these representations. Except for the third generation quarks (where this picture gets modified), the quadratic corrections to the Higgs boson mass that are generated by the Yukawas can be tolerated due to their smallness.} of $SU(5)$, a $\overline{\mathbf{5}}$, $\Psi_1$, and a $\mathbf 5$, $\Psi_2$,
\be\label{eq:fermions1}
\Psi_1=\begin{pmatrix}
\psi_1\\0\\0
\end{pmatrix}, \quad
\Psi_2=\begin{pmatrix}
0\\0\\ \psi_2
\end{pmatrix}.
\ee

As we explained in the previous section both a $\mathbf 5$ and a $\overline{\mathbf{5}}$ are needed as a consequence of the fact that in a suitable basis the automorphism $\mathcal A$ is equivalent to complex conjugation. In the Lie algebra basis we are employing, \ref{gen}, $T$-parity acts on $\Psi_{1,2}$ as
\be
T:\qquad \Psi_1\rightarrow - \Sigma_0\Omega \Psi_2
\ee

The two doublets $\psi_1$ and $\psi_2$ combine into a $T$-even and a $T$-odd combinations $\psi_1\mp\psi_2$. The even combination is identified with the SM fermion doublet and new states are required to marry the T-odd fermions in order to give them a mass. 

One may wonder how general is the fact that elementary doublets always come in couples and whether this is just a consequence of the choice to embed one of these doublets into a $\mathbf 5$ of $SU(5)$ so that $T$-parity demands a $\overline{\mathbf{5}}$ to come along. Indeed no doubling of the $SU(5)$ representation would be needed for an adjoint ($\mathbf{24}$), being this real. Nevertheless putting a single doublet inside a $\mathbf{24}$ would break $T$-parity. This picture is very general and can be understood as follows.

Collective breaking in the gauge sector together with $T$-parity invariance requires the fields to appear in the theory as complete representations of $SU(2)_1\times SU(2)_2\rtimes Z_2$ (we don't show the $U(1)$'s). $Z_2$, represented as the automorphism $\mathcal A$, acts on $SU(2)_1\times SU(2)_2$ exchanging $SU(2)_1$ and $SU(2)_2$. Thus in a viable theory only representations of the form $(\mathbf r_1, \mathbf r_2)\oplus (\mathbf r_2, \mathbf r_1)$ or $(\mathbf r, \mathbf r)$ can appear. Once these representations are reduced w.r.t. the electroweak group $SU(2)_W=SU(2)_{1+2}$ they will obviously contain an even number of doublets.

The use of elementary representations in order to decouple T-odd fermions is therefore excluded: as far as we add an even number of $SU(2)_W$ multiplets there will be one of them left out.
The only way to evade this constraint is by using composite fermions to give mass to the extra states since representations of $\mathcal H$ can be eigenstates of $T$-parity. We now show that even this possibility must be discarded. 

In \cite{tparity2, perel}, for instance, a complete representation $\psi_C$ of $SO(5)$ is introduced,
\be\label{eq:composite fermion in SU5}
\psi_C=\begin{pmatrix}
\psi_{C1} \\
\chi_C\\
\psi_{C2} 
\end{pmatrix},
\qquad T:\quad\Psi_C \rightarrow -\Omega\Psi_C\,,
\ee
 together with a $T$-parity invariant interaction 
\be\label{eq:mass and mixing}
\kappa f(\bar{\Psi}_1\Sigma_0\xi^\dagger\Psi_C+\bar{\Psi}_2\xi\Psi_C)
\ee
The above expression reduces to a mass term $\sqrt 2\kappa f$ for the $T$-odd combination $(\psi_1+\psi_2)/\sqrt 2$ and the doublet $\psi_{C2}$ when $\xi\rightarrow\langle\xi\rangle=\mathbb1_5$. 
Notice that additional chiral fermions have been introduced also in this case, a doublet $\psi_{C1}$ and singlet $\chi_C$, for which mass terms must be present. The discussion in the preceding section excludes the introduction of incomplete $SO(5)$ representations and all the vectorial representations of $SO(5)$ contain again an even number of doublets: the presence of a massless chiral fermion is unavoidable in the present framework.

There is a proposal which is often adopted in the literature to solve this puzzle. In \cite{tparity2, tparity3, perel} it is claimed it is possible to include in the theory a $\mathbf{4}$ of $Sp(4)\sim SO(5)$ decomposing as $\mathbf{4}\rightarrow \mathbf{2}+\mathbf 1+\mathbf1$ under $SU(2)_W$. In this way there would be only one additional doublet. 
The reasons that make this construction not possible are explained in \cite{CCWZ} and we repeat the argument in Appendix A for the sake of the reader. There we prove that working with the littlest Higgs coset we are not allowed to introduce spinorial representation consistently with the symmetry structure of the model.\newline

Another feature which is consequence of eq.\ref{eq:composite fermion in SU5} above (but shared by many models where composite representations are used) is the appearance of a mixing between the extra right-handed $T$-even singlet and the SM doublet after EWSB. In the specific setup we are using the mixing arises from \ref{eq:mass and mixing} when the interaction is expanded at first order in the Higgs doublet VEV
\be
M_D \left(\bar{\Psi}_2\xi\Psi_C+\bar{\Psi}_1\Sigma_0\xi^\dagger\Psi_C\right) \ni- M_D(v/f)\bar u_L \chi_C.
\ee
The effect is possible because of the definition of $T$-parity in \ref{eq:compositeTtransf} and \ref{eq:composite fermion in SU5} under which $\chi_C$ is $T$-even. In the case under consideration the left-handed singlet mixes with up-type fields with an angle $\theta\sim(v/f)(M_D/M_S)$, where $M_S$ is a singlet mass like in eq.\ref{eq:nonSO5inv}. This mixing affects $Z$-pole observables like $\sin\theta_{eff}$, the invisible width and the total hadronic width of a relative amount which is roughly given by $v^2/f^2$ for $M_D=M_S$. The former observables are known with an accuracy which is near or better than per mille. This constrains the appearance of singlets in the theory in order to avoid fine-tuning in the ratio $v/f$.\newline

To conclude this section we mention an alternative attempt \cite{tparity3, csaki} to evade the ``doublet doubling'' puzzle described above. To make it possible to add an odd number of fermionic doublets it is necessary for an odd number of them to be $T$-parity eigenstates. Instead of using composite representations one can tentatively enlarge the unbroken group adding an extra $SU(2)$ gauge site which is left invariant by the action of the $Z_2$ symmetry. The generators of this group cannot transform under $T$-parity and the associated gauge bosons must be $T$-even. $T$-parity eigenstates fermions can thus be introduced charged under this extra site, to marry the $T$-odd SM fermion partners providing them a mass.

In \cite{tparity3}, in particular,  two examples are presented where the littlest Higgs coset is enlarged to $SU(5)\times SO(5)/SO(5)$ or $SU(5)\times SU(5)/SO(5)$. In the extra site an $SU(2)\times U(1)$ vectorial subgroup
\be
Q^i_3=\begin{pmatrix}
\sigma^i/2 & &\\ & & \\ & & -\sigma^{i*}/2
\end{pmatrix}
\ee
is gauged through extra $T$-even vector bosons. To avoid contribution to $\hat S$ these vectors must be heavy and this is achieved with a strong gauge coupling $g_3\sim 4\pi$. Both the cosets contains more than a Goldstone doublet, two in the former, three in the latter. Collective breaking does not work. This is understood, in the first coset for instance, turning on two gauge couplings at a time: $g_1$ and $g_3$ or $g_2$ and $g_3$. At each time one realizes that a massless doublet is maintained in the spectrum, which is however not the same in the two cases. A term
\be\label{doublets}
\frac{g_1^2 g_3^2}{(4\pi)^4}\Lambda^2|H_1+2 H_2|^2 + \frac{g_2^2 g_3^2}{(4\pi)^4}\Lambda^2|H_1-2 H_2|^2
\ee
is indeed generated. The relative sign in \ref{doublets} is determined by $T$-parity being $H_{1,2}$ respectively even and odd. Taking now $g_3\sim 4\pi$, \ref{doublets} generates an $\mathcal O(g_{SM} f)$ mass for both the doublets.


\section{The minimal coset for the littlest Higgs with $T$-parity}\label{mincoset}
A consistent implementation of $T$-parity in the little-Higgs framework must take care of all the issues we discussed so far. In this section we propose such a model. We will not try the construction of a UV complete model. The interested reader is referred to \cite{csaki} for an attempt at this.

It should now be clear that the crucial aspect is the introduction of the SM fermions. We found that we need an odd number of electroweak doublets and this cannot be achieved with elementary fermions without enlarging the coset structure. Since we would like to obtain a model where the little hierarchy problem is solved with no fine-tuning, we want to avoid incomplete composite representations and also additional $T$-even gauge sites. 

To avoid custodial breaking contributions coming from the $\sigma$-model we take the littlest Higgs construction as a starting point. We will construct a model where the original coset is enlarged in such a way as to include an extra unbroken $SU(2)\times U(1)$ global factor. This factor will allow the inclusion of a single extra doublet consistently with $T$-parity. We consider the coset 
\be\label{newcoset}
\frac{SU(5)}{SO(5)}\times \frac{[SU(2)\times U(1)]_L\times  [SU(2)\times U(1)]_R}{ [SU(2)\times U(1)]_V}.
\ee
The $\sigma$-model is parametrized by two matrices $\Sigma$ and $X$ transforming as
\be
\Sigma\rightarrow g \Sigma g^T,\quad X \rightarrow U_L X V^\dagger=V X U_R^\dagger\,,
\ee
where $g$ and $U_{L(R)}$ are respectively elements of $SU(5)$ and $[SU(2)\times U(1)]_{L(R)}$ while $V$ belongs to $[SU(2)\times U(1)]_V$. We also define $ X^2$, transforming linearly as $X^2\rightarrow U_L X^2 U_R$. $\Sigma$ and $X$ are written in terms of the Goldstone bosons according to the usual exponential notation
\be
\Sigma=e^{2i \Pi_\Sigma/f}\Sigma_0\,, \qquad X=e^{i \Pi_X/f}.
\ee
The matrix $\Pi_\Sigma$ contains the 14 Goldstone bosons parametrizing the coset $SU(5)/SO(5)$ while $\Pi_X$ contains the new 4 Goldstone bosons. 

The $SU(5)$ global symmetry contains two $[SU(2)\times U(1)]_{1,2}$ subgroups associated to the generators
\be\label{generators}
Q_1^a=\begin{pmatrix}
\sigma^a/2 & 0 & 0\\
0 & 0 &0 \\
0 & 0 &0 
\end{pmatrix},\qquad Y_1=\textrm{diag}\left(\tfrac{3}{10},\tfrac{3}{10},-\tfrac{2}{10},-\tfrac{2}{10},-\tfrac{2}{10}\right)
\ee
\ben
Q_2^a=\begin{pmatrix}
0 & 0 & 0\\
0 & 0 &0 \\
0 & 0 & -\sigma^{a*}/2 
\end{pmatrix},\qquad Y_2=\textrm{diag}\left(\tfrac{2}{10},\tfrac{2}{10},\tfrac{2}{10},-\tfrac{3}{10},-\tfrac{3}{10}\right)
\een
We gauge the two combinations $[SU(2)\times U(1)]_{1+L}$ and  $[SU(2)\times U(1)]_{2+R}$. 
At the scale\footnote{In principle the symmetry breaking scale can be different but we take them equal for sake of simplicity.} $f$ the global symmetry is broken by the VEVs
\be
\left<\Sigma\right>=\Sigma_0,\qquad \left<X^2\right>=\mathbb1_2,
\ee
in such a way that the electroweak group is identified with the vectorial subgroup $[SU(2)\times U(1)]_{1+2+L+R}$.
The scalars are thus decomposed according to their electroweak quantum numbers
\ba
\Pi_\Sigma:&& \mathbf 3_C\oplus\mathbf 3_R\oplus\mathbf 2_C\oplus\mathbf 1_R\\
\Pi_X:&& \mathbf 3_R\oplus\mathbf 1_R
\ea
where the labels $C$, $R$ indicate respectively a complex or a real representation. In more details we write:
\be\label{goldstoni}
\Pi_\Sigma=\begin{pmatrix}
 \frac{\tau \cdot \sigma}{2} +\frac{\phi_0 }{2\sqrt 5} &   H & T\\ H^\dagger & -\frac{2}{\sqrt 5}\phi_0 &  H^T\\
T^\dagger &  H^* &  \frac{\tau \cdot \sigma^*}{2}+\frac{\phi_0}{2\sqrt 5} 
\end{pmatrix}, \qquad \Pi_X=\frac{\pi\cdot \sigma+\pi_0\mathbb 1_2}{2} 
\ee
where $T$ is a complex triplet with unit hypercharge and $H$ is the Higgs doublet:
\be
T=\begin{pmatrix} T^{++} & T^+/\sqrt2\\ T^+/\sqrt2&  T^0 \end{pmatrix},\qquad H=\frac1{\sqrt2}\begin{pmatrix}h_1+i h_2\\ h_3+ih_4 \end{pmatrix},\qquad \left<H\right>=\frac{1}{\sqrt 2}\begin{pmatrix}0\\v \end{pmatrix}
\ee

According to the discussion of sec. 2 the action of $T$-parity can be defined as the combination of the automorphism $\mathcal A$ that changes sign to the broken generators together with an $SU(2)_W$ rotation:
\be
\Sigma\rightarrow  \Omega\Sigma_0 \Sigma^\dag \Sigma_0\Omega\,, \qquad X\rightarrow X^\dagger\,.
\ee
The above choice fixes the parity assignments of the goldstone bosons
\ba
\Pi_\Sigma:&& \mathbf 3^-_C\oplus\mathbf 3^-_R\oplus\mathbf 2^+_C\oplus\mathbf 1^-_R\\
\Pi_X:&& \mathbf 3^-_R\oplus\mathbf 1^-_R
\ea
The lagrangian for the $\sigma$-model is
\be
\mathcal L=\frac{f^2}8\text{Tr}[D\Sigma^* D\Sigma]+ \frac{f^2}{4}\text{Tr}[D{X^2}^\dagger DX^2],
\ee
The covariant derivatives are written explicitly in Appendix B. Here we limit to note that the breaking of the global group results in a set of heavy $T$-odd vectors 
\be
B_H=\frac{B_1-B_2}{\sqrt 2},\qquad W^i_H=\frac{W^i_1-W^i_2}{\sqrt 2},
\ee
orthogonal to the $SM$ gauge bosons, whose masses are 
\be
m_{B_H}= \frac{\sqrt 6}{5} g'f, \qquad m_{W_H}= \sqrt2g f.
\ee
at lowest order in $v/f$. In the unitary gauge the following four combinations of scalars are removed from the spectrum:
\be
G_0= \sqrt{\frac 5 6}\phi_0- \sqrt{\frac 1 6}\pi_0,\qquad G_i=\frac{\pi_i+\tau_i}{\sqrt 2}.
\ee
The two new physical $T$-odd Goldstone bosons, an $SU(2)$ triplet and a singlet, both with vanishing hypercharge, are identified with the combinations
\be
s= \sqrt{\frac 1 6}\phi_0+ \sqrt{\frac 5 6}\pi_0,\qquad \varphi_i=\frac{-\pi_i+\tau_i}{\sqrt 2}.
\ee


\subsection{Fermionic sector}
Let us analyze how the modified coset allows us to reproduce the SM low energy spectrum avoiding the difficulties discussed in section 2. We begin considering light fermions (all the leptons and all the quarks except the third generation).
Two left-handed doublets, $\psi_1$ and $\psi_2$, are introduced for each one in the SM. They are chosen to transform under $\left[SU(2)\times U(1)\right]_{1+L}$ and $\left[SU(2)\times U(1)\right]_{2+R}$ respectively. Together with them an additional composite right-handed doublet $\psi_C$ charged under $\left[SU(2)\times U(1)\right]_{V}$ and two right-handed singlets, $u_R$ and $d_R$, are considered. Their $T$-parity transformation are fixed to be
\be
\psi_1\leftrightarrow -\psi_2,\qquad \psi_C\leftrightarrow -\psi_C,\qquad u_R \leftrightarrow u_R,\qquad d_R \leftrightarrow d_R.
\ee
We add to the Lagrangian the following term
\be\label{eq:oddcombination}
\frac{\kappa f}{\sqrt 2}\left(\overline\psi_1  \sigma_2 X\Psi_C +\overline \psi_2\sigma_2 X^\dag\Psi_C\right)
\ee
which contains, at the zeroth order in the Goldstone fields, a mass term for the $T$-odd combination $(\psi_1+\psi_2)/\sqrt 2=\sigma_2(U,~ D)^T$, $m_{Q^-}=\kappa f$. 
Notice that the term in \ref{eq:oddcombination} preserves the whole global symmetry. This should be compared to the analog term in \cite{perel} and already mentioned above (see eq. \ref{eq:mass and mixing}), which alone already breaks the global $SU(5)$ to an $SU(3)\times SU(2)$ subgroup. 

\begin{figure}[t] 
\centering%
\includegraphics[scale=1]{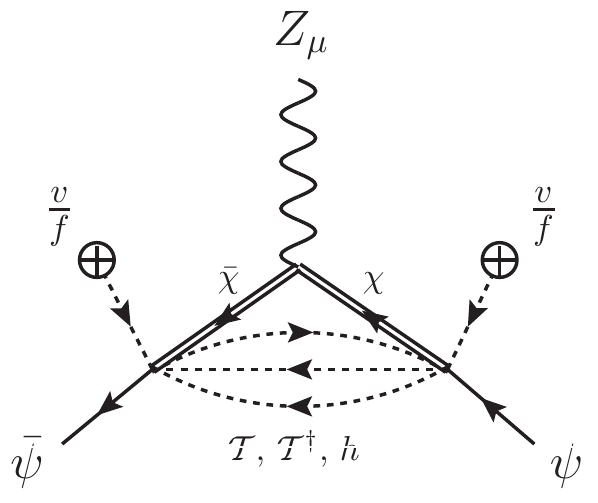}
\caption{\small Example of a Feynman diagram generating the vertex correction discussed in eq.\ref{2loopvertexcorrectionseq}.} \label{2loopvertexcorrectionsfig}
\end{figure}

As anticipated in the previous section and explained in \cite{perel} the mass $m_{Q^-}$ is what provides the cutoff to the quadratic divergence in the four-fermions operators $\mathcal O_{J_1J_2}$ shown in eq. \ref{O4F}. It is thus important for $\kappa$ to be small enough. In \cite{perel} the lightness of these $T$-odd fermions and thus the smallness of $\kappa$ have however extra responsibilities\footnote{We highlight these differences, even comparing with a model with major phenomenological flaws, to illustrate various points which must be kept in mind for further model building.}. In first place, there are vertex corrections to the light SM fermions couplings to gauge bosons coming for instance from the diagram in Fig.\ref{2loopvertexcorrectionsfig}. These are estimated to be of order

\be\label{2loopvertexcorrectionseq}
\frac{\delta g}{g} \approx \frac{\kappa^2}{16\pi^2}\frac{v^2}{f^2},
\ee
thus requiring a perturbative value for $\kappa$. Such corrections are further suppressed in the setup presented in eq. \ref{eq:oddcombination}: if all the other weak couplings are turned off there is no connection between the two $\sigma$-models parametrized by $X$ and $\Sigma$. Second, after electroweak symmetry breaking the term $\overline \psi^- HH^\dagger \psi_{C2}$ contained in the expansion of eq. \ref{eq:mass and mixing} lifts the degeneracy among the masses of the up and down component of the $T$-odd doublet introducing a contribution to the $\hat T$ parameter given by
\be
\Delta \hat T=\frac{N_c\kappa^2}{384\pi^2}\frac{v^2}{f^2}
\ee
which is enhanced by the large number of doublets in the SM (three uncolored and three colored ones). These effects are again naturally suppressed in our implementation since the $T$-odd doublets stay degenerate in the gaugeless limit.

Gauge invariance w.r.t. the various $U(1)$ factors is achieved with a suitable choice of the hypercharges. We explain this choice together with the way the Yukawa couplings of light fermions are introduced in Appendix B. As usual in little Higgs model such Yukawa couplings are not asked to preserve any shift symmetry of the Higgs boson, since the breaking they induce is very small and not relevant for naturalness.
\newline

The role of the enlarged coset should now be clear. It is the minimal extension which allows for the inclusion of a single doublet which is an eigenstate of $T$-parity, avoiding composite incomplete representations. This is the minimal requirement to write a mass term for the heavy fermions. We also notice that no singlet accompanies the doublet nor new $T$-even gauge bosons, evading the difficulties mentioned at the end of section 2. Other pleasing characteristics of this new setup have been observed in the last paragraph. We stress that, as showed in Appendix B, the enlarged coset must contain a broken $U(1)$ factor in such a way as to achieve gauge invariance w.r.t. hypercharge.\newline

The top quark requires special attention since in this case the collective breaking mechanism must be implemented also for the Yukawa interactions. To the fermionic content employed to reproduce the light SM fermions we add two left-handed singlets, $\chi_1$ and $\chi_2$, and two right-handed colored singlets, $\tau_1$ and $\tau_2$. Their transformation properties under $T$-parity are fixed requiring
\be
\chi_1\leftrightarrow \chi_2,\qquad \tau_{1,2}\leftrightarrow -\tau_{1,2}.
\ee
Again, we add the interaction required to give mass to the $T$-odd combination of doublets 
\be\label{massoddtop}
\frac{\kappa f}{\sqrt 2} \left(\overline{\psi}_1 \sigma_2 X{\tau}_C+ \overline {\psi}_2\sigma_2X^\dagger{\tau}_C\right)
\ee
where $\tau_C$ is the analog of $\psi_C$ for the top quark. To write the top Yukawa in a way which is manifestly consistent with collective breaking, we embed the two doublets $\psi_{1,2}$ and the two singlets $\chi_{1,2}$ in incomplete representations of $SU(5)$, a $\mathbf 5^*$ and a $\mathbf 5$
\be
\mathcal T_1=\begin{pmatrix} \psi_1\\ \chi_1\\0
\end{pmatrix},\qquad
\mathcal T_2=\begin{pmatrix} 0\\\chi_2\\\psi_2
\end{pmatrix}
\ee
in such a way that $T$-parity acts as
\be
\mathcal T_1\leftrightarrow -\Omega\Sigma_0\mathcal T_2.
\ee
The Yukawa coupling is then as follows \cite{littlest,tparity3,perel}
\be\label{yukawatop}
\frac{\lambda_1 f}{4}\epsilon_{ij}\epsilon_{abc}\left(\Sigma_{ai}\Sigma_{bj}\overline{\mathcal T}_{1c}  - \widetilde\Sigma_{ai} \widetilde\Sigma_{bj}(\overline{\mathcal  T}_2\Sigma_0\Omega)_c \right)t+\frac{\lambda_2 f}{\sqrt2}(\overline \chi_1\tau_1+\overline \chi_2\tau_2)
\ee

All the terms appearing in the previous equations are multiplied by weak, $\mathcal O$(1), couplings. 
As discussed above, eq. \ref{massoddtop} preserves the whole global symmetry in eq. \ref{newcoset}. The same is trivially true for the coupling $\lambda_2$. Moreover both terms inside the first round parentheses in eq. \ref{yukawatop} are individually invariant under two different, global, $SU(3)$ symmetries under which the Higgs doublet transforms non-linearly.
This, together with a set of spurionic $U(1)$ symmetries than can be assigned to the various couplings appearing in \ref{massoddtop} and \ref{yukawatop} (see next section), implies the lowest order UV contribution to the Higgs boson mass to have at most the magnitude
\be
\Delta m_h^2\approx \frac{|\lambda_1|^2|\lambda_2|^2}{(16\pi^2)^2}\Lambda^2,
\ee
which causes no naturalness problem. 
Defining the $T$-parity eigenstates
\be
\frac{\psi_1-\psi_2}{\sqrt 2}=\sigma_2 q= \sigma_2\begin{pmatrix} t_L\\ b_L\end{pmatrix}\,\qquad \chi^\pm = \frac{\chi_1\pm\chi_2}{\sqrt 2}\,\qquad  \tau^\pm = \frac{\tau_1\pm\tau_2}{\sqrt 2}.
\ee
we can expand $\Sigma$ and $X$  at all orders in the ratio $v/f$ and at zeroth order in the other Goldstone fields fields to get
\be
-\frac{\lambda_1f}{\sqrt 2} \left(\frac{s_v}{\sqrt 2}\right)\overline t_L t_R+\frac{\lambda_1 f}{\sqrt 2}\left(\frac{c_v + 1}{2}\right)\bar \chi^+ t_R+\frac{\lambda_2 f}{\sqrt2}\bar \chi^+\tau^++ \frac{\lambda_2 f}{\sqrt2} \bar \chi^-\tau^-+\kappa f (\overline T_L T_R +\overline B_L B_R)
\ee
where $s_v\equiv \sin\sqrt 2v/f$, $c_v=\cos\sqrt 2 v/f$ and $T_R$, $B_R$, are the two components of the composite doublet $\mathcal T_C$.
At leading order in $v/f$ the right-handed top quark is thus identified with the massless combination $(\lambda_2 t-\lambda_1\tau^+)/\sqrt{\lambda_1^2+\lambda_2^2}$, and the top Yukawa coupling is $\lambda_t=\lambda_1\lambda_2/\sqrt{\lambda_1^2+\lambda_2^2}$. 
After EWSB the top sector is composed by four Dirac fermions: a $T$-odd singlet of mass $m_{T^-}=\lambda_2 f/\sqrt 2$, a degenerate $T$-odd doublet of mass $\kappa f$ and two $T$-even states. These latter are identified with the top quark itself and a heavy partner. Their masses are (working at leading order in $v/f$)
\be\label{toppartners}
m_t=\frac{\lambda_1\lambda_2 }{\sqrt{\lambda_1^2+\lambda_2^2}}\frac{v}{\sqrt 2},\qquad m_{T^+}=\sqrt{\lambda_1^2+\lambda_2^2}\frac{f}{\sqrt 2}.
\ee
The presence of a singlet mixing with the left-handed top quark is unavoidable for collective breaking. The resulting modification in the top quark couplings to electroweak gauge bosons place no relevant constraint on this mixing, since these are not measured at the present time.

\subsection{Collective breaking and scalar potential}

The structure of the scalar potential in the present model is necessarily more involved than the one in the original littlest Higgs owing to the presence of the extra goldstones. 

Two necessary requirements for a viable little Higgs model are the generation of a $\mathcal O(g^2_{SM})$ strength quartic coupling and the absence of contributions to the Higgs boson mass up to two loops\footnote{A note about the terminology. We call an $n$-loop contribution to the coefficient of an operator, the NDA size of $n$ weakly coupled loops
\be\label{ndasize}
f^2\Lambda^2\left(\frac{g_{SM}^2}{16\pi^2}\right)^n\mathcal O\left(\frac{\pi}{f},\frac{\partial}{\Lambda},\ldots\right)
\ee
with $g_{SM}$ a weak coupling as those in the SM. Eq. \ref{ndasize} neglect possible logarithmic enhancement.
}.
These are non trivial requests since the explicit breaking of the Higgs' shift symmetry required for the generation of its quartic coupling must not feed into an Higgs boson mass at the same order. In \cite{wacker} a general analysis of the cosets where this can be achieved is performed. We will discuss these issues in our modified version of the little Higgs.
\newline

There are two sources of explicit breaking of the global symmetry, the gauge and the Yukawa couplings. The low energy scalar potential is thus expected to contain certain symmetry violating operators. Their coupling structure and size are determined by the analogous terms contained in the Coleman-Weinberg (CW) potential generated by loops of gauge bosons and fermions. 

We start by examining what happens to the coset turning on the various couplings. A necessary requirement for collective breaking is that each coupling, singularly, must preserve enough of the global symmetry as to leave in the spectrum a massless doublet with the quantum numbers of the Higgs doublet.

Starting with the gauge interactions and turning on, for instance, the $SU(2)_2\times U(1)_2$ gauge copy the residual global symmetry is (indicating also the gauge symmetries among the global ones)
\be\label{gbreak}
\frac{SU(3)_1\times \left[SU(2)\times U(1)\right]_{2+R}\times \left[SU(2)\times U(1)\right]_{L}\times U(1)_{R}}{[SU(2)\times U(1)]_{1+2+L+R}\times U(1)_{L+R}}
\ee
where $SU(3)_{1(2)}$ is the upper-left (lower-right) $SU(3)$ subgroup containing the $SU(2)_{1(2)}$ generators. At this level only the complex triplet $T$ develop a potential. Turning on also the second gauge coupling the coset is replaced by
\be
\frac{\left[SU(2)\times  U(1)\right]_{1+L}\times \left[SU(2)\times U(1)\right]_{2+R}\times U(1)_{L}\times U(1)_{R}}{[SU(2)\times U(1)]_{1+2+L+R}\times U(1)_{L+R}}.
\ee
This contains the real triplet and the real singlet eaten by the $T$-odd combination of gauge vectors, together with a real scalar singlet which, at this point, is an exact Goldstone boson. For what concerns the Higgs doublet one thus expects UV contributions to its potential, which are analytic in the gauge coupling constants, to originate at this order. This ensures the absence of quadratically divergent analytic contribution to the Higgs mass from the gauge sector.

We stress that these arguments ensures the absence of UV contributions to the mass of the Higgs boson but not those of infrared (IR) contributions. The formers are analytic in the parameters, while the latters can involve all the couplings (thus breaking all the symmetries) in a non-analytic way. We will come back to this point later discussing the generation of the Higgs quartic coupling which is an effect of the second kind.

A similar reasoning proceeds with the coupling of the fermions. For clarity we distinguish among the equal coefficients of $T$-parity conjugated terms using a tilde for one of them, since the symmetries broken by each of the two are different (eg. we call $\lambda_1$ the first term in the first round parentheses in eq. \ref{yukawatop} and $\widetilde \lambda_2$ the second.). We already sketched the symmetry breaking pattern in the section devoted to the top lagrangian. The $\kappa$ and $\widetilde\kappa$ couplings alone preserve the whole global symmetry in eq. \ref{newcoset}. The same is true for $\lambda_2$ or $\widetilde\lambda_2$.  On the other hand $\lambda_1$ breaks the coset to
\be\label{l1break}
\frac{SU(3)_1\times \left[SU(2)\times U(1)\right]_2}{[SU(2)\times U(1)]_{1+2}}\times \frac{[SU(2)\times U(1)]_L\times[SU(2)\times U(1)]_R}{ [SU(2)\times U(1)]_V}
\ee
thus allowing the generation of a one-loop potential for the complex triplet $T$. The breaking induced by $\widetilde \lambda_1$ is obtained from eq. \ref{l1break} with the replacement $1\leftrightarrow 2$. The breaking induced by $\lambda_1$ and $\widetilde\lambda_1$ simultaneously preserves just the gauge symmetries of the original littlest Higgs, so in the present case
\be
\frac{[SU(2)\times U(1)]_1\times[SU(2)\times U(1)]_2}{ [SU(2)\times U(1)]_{1+2}}\times\frac{[SU(2)\times U(1)]_L\times[SU(2)\times U(1)]_R}{ [SU(2)\times U(1)]_V}
\ee
allowing a UV potential for the Higgs boson. The necessary UV breaking of the Goldstone symmetries associated with the extra fields we introduced, the triplet $\pi_i$ and the singlet $\pi_0$, are allowed when all $\lambda_1$, $\widetilde\lambda_1$, $\kappa$ and $\widetilde\kappa$ are considered.

Notice that unlike the gauge case, eq. \ref{l1break}, which shows the smallest global symmetry preserved when just one coupling among fermions is turned on, is still not enough to guarantee the absence of one-loop UV contributions to the Higgs potential. Indeed even if it is true that a quadratically divergent contribution to its mass can be at most quadratic in the weak couplings, we have up to now no motivation to believe that a term proportional to $\lambda_1^\dagger\widetilde\lambda_1$ cannot appear in a one-loop diagram. That this is not the case can be ascertained assigning spurionic $U(1)$ charges to couplings and fermion fields. Consider the assignments in Table \ref{U1spurions}.

\begin{table}[h]
\centering
\begin{tabular}{cccccccccccc}
\hline
$\mathcal T_1$& $\mathcal T_2$& $\mathcal T_C$ & $t$ & $\tau_1$& $\tau_2$&$\kappa$&$\widetilde \kappa$&$\lambda_1$&$\widetilde \lambda_1$&$\lambda_2$&$\widetilde \lambda_2$\\
\hline
$a$& $b$& $c$ & $d$ & $e$& $f$&$a-c$&$b-c$&$a-d$&$b-d$&$a-e$&$b-e$\\
\hline
\end{tabular}
\caption{\small Spurionic $U(1)$ charge assignments. No quadratic combination except the trivial ones $\kappa^\dagger\kappa$, $\lambda^\dagger\lambda$ is allowed to appear.}
\label{U1spurions}
\end{table}

No quadratic term apart from the trivial ones $\kappa^\dagger\kappa$, $\lambda^\dagger\lambda$, is allowed. UV contributions to the Higgs boson potential must go through at least four coupling insertions for instance, $\lambda_1^\dagger\widetilde\lambda_1\lambda_2\widetilde\lambda_2^\dagger$ or $\lambda_1^\dagger\widetilde\lambda_1\kappa\widetilde\kappa^\dagger$. The last term we wrote will in particular appear in the contributions to the UV potential for the extra singlet and triplet. So far we showed that what we can call ``UV collective breaking'', operates for the Higgs doublet as well as for the new scalars.
\newline

As emphasized above we still have to make sure that the IR contribution to the Higgs potential do not spoil the good feature of its UV counterpart. The discussion follows those already present in the existing literature \cite{littlest, dangsinglets, wacker}. 
The quadratically divergent part of the scalar potential is given by
\be\label{CWquad}
\Delta V_{CW}= c_V f^2 \tr[M^2_V(\Sigma, X)]+c_F f^2 \tr[M_f M_f^\dagger(\Sigma, X)],
\ee 
where $M_V$ and $M_f$ are respectively the vector boson and fermion mass matrices in the background field of $\Sigma$ and $X$. The appearance of quadratically divergent operators suggests the presence of tree level counterterms with the same symmetry structure of \ref{CWquad}: the coefficient $c_V$, $c_F$ have thus $\mathcal O(1)$ indetermination.

From the discussion in the previous paragraph we already know that turning on a single coupling among $g_1$, $g_2$, $\lambda_1$, $\widetilde\lambda_1$ only the complex triplet can obtain a potential. Expanding eq. \ref{CWquad} at lowest order in the Goldstone fields one gets
\be\label{CWquad1}
\Delta V_{CW}= c g_{SM}^2 f^2\left| T_{ij}+\frac{i}{2f}(H_iH_j+H_jH_i)+\ldots \right|^2+\widetilde c \widetilde g_{SM}^2 f^2\left| T_{ij}-\frac{i}{2f}(H_iH_j+H_jH_i)+\ldots\right|^2
\ee
with $g_{SM}$ a generic SM coupling either gauge or Yukawa and where by $T$-parity $g_{SM}\equiv \widetilde g_{SM}$, $c=\widetilde c$ and $T\rightarrow -T$ so that the above term is invariant. If only one of the two terms in \ref{CWquad1} is present one can use a transformation in the cosets \ref{gbreak}, \ref{l1break} to end up with a potential containing just the complex triplet, in particular a mass term $\mathcal O(g_{SM}f)$. In this case the low energy quartic coupling vanishes when we sum the contribution from the expansion of the square in \ref{CWquad1} and the one obtained after integrating out the complex triplet. In a similar way, all the terms involving other goldstones, which we implied by the dots, are unphysical. 

When eq. \ref{CWquad1} is considered as a whole the potential for the other goldstones can no longer be eliminated with a global rotation and a quartic coupling for the Higgs is generated. In a theory without $T$-parity this effects is given by
\be
\Delta \mathcal L=\left[\frac{(c g^2_{SM}- \widetilde c\widetilde g_{SM}^2)^2}{c g^2_{SM}+ \widetilde c\widetilde g_{SM}^2}-(c g^2_{SM}+ \widetilde c\widetilde g_{SM}^2)\right] (H^\dagger H)^2.
\ee
The first term in the square parentheses is obtained in the low energy lagrangian after integrating out the heavy complex triplet, while the second term is just the sum of the two quartic couplings coming from eq. \ref{CWquad1}. In a theory with $T$-parity the first contribution is absent since the coupling $THH$ is $T$-odd. 

This structure can be better understood as follows. Let us consider a $\sigma$-model based on the coset $\mathcal G/\mathcal H$. Some couplings can be turned on in such a way as to reduce the symmetry structure to $\mathcal G'/\mathcal H'$. Loops containing insertions of these couplings will thus generate a potential for the goldstones, $\pi_H$, whose generators have been explicitly broken. Working in the ``$\mathcal G/\mathcal H$ basis'', that is with the goldstones parametrizing the $\mathcal G/\mathcal H$ coset, the potential will generically contain all the Goldstone fields. This is so because, tipically, the broken generators in $\mathcal G'/\mathcal H'$ are combinations of broken and unbroken generators in $\mathcal G/\mathcal H$. There must exist, however, a transformation in $\mathcal G'$ such that
\be\label{VpiH}
V(\pi_L, \pi_H)\rightarrow V'(\pi'_H),
\ee
with the new potential depending only on the ``were-goldstones''.

The crucial feature of the littlest Higgs coset is thus the fact that the mediator of the quartic is a complex triplet and not, for instance, a real scalar \cite{dangsinglets}. This implies that the potential generated at one loop is a gauge invariant function of the complex triplet only: $V'(T')=V(T+\ldots)$. Hence the expansion of eq. \ref{CWquad1} cannot contain, just by gauge invariance, an order $g_{SM} f$ mass term for the Higgs nor such a mass term can be generated integrating out the complex triplet.

Keeping terms with more than two fields (but just $H$ or $T$) in the expansion of \ref{CWquad1} one encounters structures like $T_{ik}H^\dagger_kH_j$; expanding the square these will in turn generate an operator
\be\label{tripletsplitting}
g^2_{SM}H^\dagger T T^\dagger H
\ee
which, after EWSB, splits the squared masses of the components of the triplet of an amount $\mathcal O(g^2_{SM}v^2)$. It is important to keep in mind that closing a $TT^\dagger$ loop in \ref{tripletsplitting} will not generate a $\mathcal O (g_{SM}f)$ mass for the Higgs boson. Loops of goldstones do not bring in new symmetry breaking effects so this mass would be an UV contribution which, as we already know, is not allowed by gauge invariance to appear in a generic potential of the form \ref{VpiH}.

Similar considerations are valid for terms involving the new goldstones $\varphi$ and $s$. One can enumerate all the possible structures with the same quantum numbers of the triplet $T$, that can in principle appear in the potential $V(T,\pi_L)$ obtained from a gauge invariant potential $V'(T')$ after a generic transformation in \ref{gbreak}, \ref{l1break}. Since both $s$ and $\varphi$ have vanishing hypercharges, terms not involving the complex triplet $T$ will contain at least a power of $(H^\dagger H)^2$. This automatically ensures the absence of one-loop masses for the extra goldstones but also the absence of one-loop contributions to operators like
\be\label{dangoperators}
H^\dagger H s^2,\quad s\varphi^i H^\dagger\sigma^i H,\quad (\varphi^i H^\dagger \sigma^i H)^2.
\ee
This is a relevant phenomenological feature as we will explain in the following.
\newline

Part of the UV effects beaking the whole global symmetry are contained in the logarithmically divergent CW potential 
\be\label{CWlog}
\Delta V_{CW}=\frac{3}{64\pi^2}\tr\left[M_V^4 \log\frac{M_V^2}{\Lambda^2}\right]-\frac{3}{16\pi^2}\tr\left[M_f M_f^\dagger M_f M_f^\dagger\log \frac{M_f M_f^\dagger}{\Lambda^2}\right]
\ee
respectively from gauge bosons and fermions. Inspecting eq. \ref{CWlog} one finds a mass for the Higgs boson doublet coming both from the gauge and from the fermionic part. The gauge part gives also a contribution to the mass of the real triplet $\varphi$. No contribution to the mass of the scalar singlet $s$ is present in \ref{CWlog}. This is not a problem since we are not including loops of goldstone bosons. Being the global symmetry completely broken they will introduce all possible symmetry breaking operators. Eq. \ref{CWlog} contains for instance all the operators mentioned in eq. \ref{dangoperators}. It is enough to close a $HH^\dagger$ loop in the first of them to get the desired mass term for $s$.
\newline

The previous considerations imply that all the light scalars share roughly the same mass which is a two-loop effect
\be\label{scalars}
m^2_H\approx m^2_\varphi \approx m^2_s\approx \frac{g_{SM}^4}{16\pi^2}f^2.
\ee
The sizes of the operators in eq. \ref{dangoperators} are also two-loop. The last two in particular, which are responsible after EWSB for the mass splitting among the charged and neutral component of the triplet $\phi$,  are generated with coefficients
\be\label{realtripletsplit}
\frac{g_{SM}^4}{16\pi^2}s \varphi^i H^\dagger\sigma^i H,\qquad \frac{g_{SM}^4}{16\pi^2 f^2}(\varphi^i H^\dagger\sigma^i H)^2.
\ee

\subsection{Phenomenology}\label{phenomenology}

In the two previous sections we introduced a model which realizes in a minimal way the ideas of $T$-parity in the littlest Higgs model. There are three aspects we think are worth a brief discussion. These are the existing bounds on the model from precision electroweak observables, the possible implications for cosmology and the potential for discovery at future colliders. \newline

$T$-parity relieves the tension with precision observables found in plain little-Higgs models, prohibiting corrections coming from tree-level exchange of the new states. The lowest order effects come at one-loop and are thereby small. 
One thus expects our model to be only mildly constrained by EWPT. The most important effects are encoded in the modification of the $\hat T$ parameter, in particular due to the enlarged top sector. The top quark is an admixture of a doublet and a singlet and this results in a contribution to $\hat T$ given by
\be\label{Ttop}
\Delta \hat T=\hat T_{SM}\left(-\frac{2 c_L^2}{1-x_t}\log x_t-2 +s_L^2+\frac{s_L^2}{x_t}\right),\qquad x_t=\frac{m_t^2}{m_T^2}
\ee
\begin{equation*}
\hat T_{SM}=\frac{3 G_F m_t^2}{8\sqrt 2\pi^2}\approx 0.009.
\end{equation*}
where $\hat T_{SM}$ is the custodial breaking contribution of the top quark in the SM, and $s_L\equiv \sin \theta_L$, $c_L\equiv \cos \theta_L$ rotates the interaction eigenstates to the mass eigenstates
\be
\begin{pmatrix}
t_L\\ \chi^+
\end{pmatrix}
=
\begin{pmatrix}
\cos\theta_L& \sin\theta_L\\ -\sin\theta_L&\cos\theta_L
\end{pmatrix}\begin{pmatrix}
\widetilde t_L\\ \widetilde T_L
\end{pmatrix}.
\ee
Working at leading order in $x_t$, eq. \ref{Ttop} reduces to
\be\label{Ttopapp}
\frac{\Delta \hat T}{\hat T_{SM}}=2\left(\frac{\lambda_1}{\lambda_2}\right)^2\frac{m_T^2}{m_t^2}\left[\log \frac{m_t^2}{m_T^2}-1+\left(\frac{\lambda_1}{\lambda_2}\right)^2\right],\qquad \frac{\lambda_1}{\lambda_2}\frac{m_t}{m_T}\approx s_L
\ee
This contribution, being positive, can help accommodating a somewhat heavier Higgs boson with respect to the typical SM scenario.
The same structure of eq. \ref{Ttopapp} is shared by the correction to the $Zb_Lb_L$ vertex\footnote{The correction is defined starting from the tree level $Z b_Lb_L$ vertex
\be
\left(-1/2+s_W^2/3+\delta g_{Zb_Lb_L}\right)\overline b_L\gamma^\mu b_L Z_\mu.
\ee}, when calculated in the same $x_t\ll1$ approximation and in the ``gaugeless'' limit of the SM 
\cite{barbierigaugeless}, $m_t,m_T\gg m_W$.
\be
\frac{\delta g_{Zb_Lb_L}}{\delta g^{SM}_{Zb_Lb_L}}\approx\frac{\Delta \hat T}{\hat T_{SM}},\qquad \delta g_{Z b_L b_L}^{SM}=\frac{ G_F m_t^2}{8\sqrt 2\pi^2}\approx 0.003.
\ee
Another subleading contribution to $\hat T$ comes from loops of the heavy vectors $W_H$. These are splitted after EWSB by an amount
\be
\Delta M_H^2\equiv M^2_3-M^2_\pm=\frac1 2 g^2 f^2 \sin^4\frac{v}{\sqrt 2 f}
\ee
where we neglected terms proportional to $g'$. This contribution is evaluated in \cite{tparity2, perel} and is found to be
\be\label{tvectors}
\Delta T_V=-\frac{G_F\Delta M_H^2}{8\sqrt 2\pi^2}\log\frac{\Lambda^2}{M_H^2};
\ee
Being \ref{tvectors} logarithmically divergent, a tree level counterterm is expected in the low energy lagrangian. We assume this term to be small in such a way to neglect it. This is justified by the fact that the NDA size of such counterterm will be the same of eq. \ref{tvectors} but without the logarithmic enhancement. The piece in eq. \ref{tvectors} is parametrically subdominant w.r.t. the top sector contribution, but gives the strongest constraint when $\lambda_1/\lambda_2$ is small. 

A further violation of the custodial symmetry comes from loops of the scalar triplets $T$ and $\varphi$ since the component of both are splitted after EWSB. Given the natural size of these mass splittings from eq. \ref{tripletsplitting}, \ref{realtripletsplit}, their contribution to $\hat T$ is negligible.

\begin{figure}[t] 
{\includegraphics[scale=0.9]{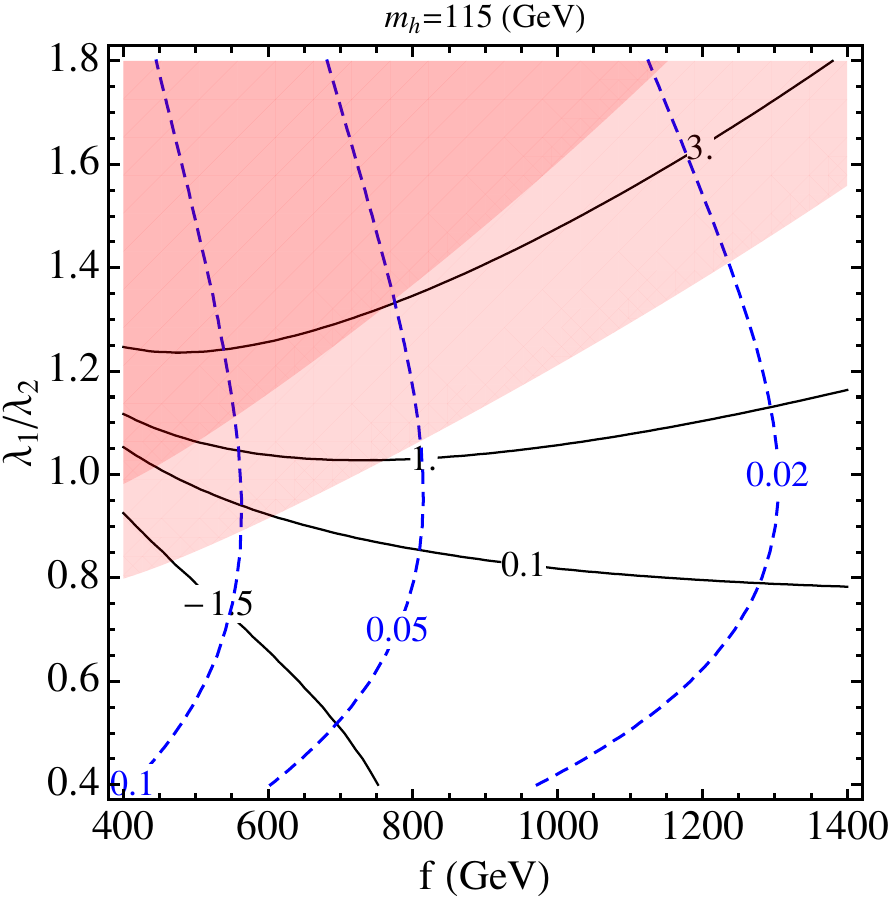}\includegraphics[scale=0.9]{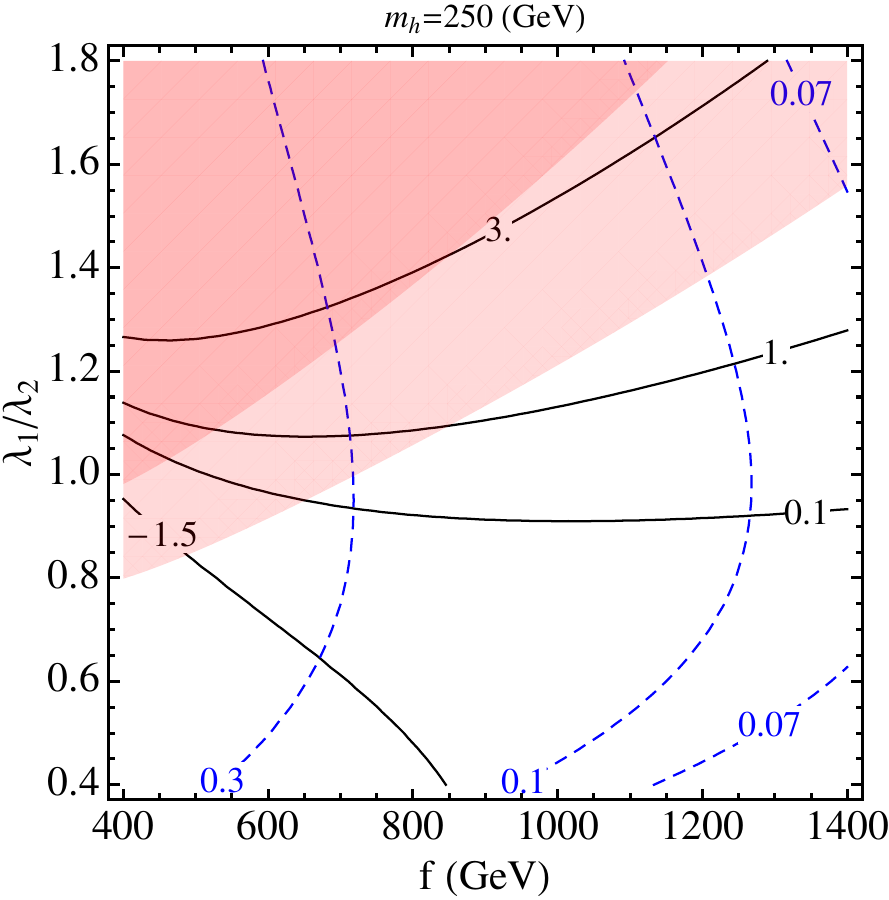}}
\caption{\small Constraints on the parameter space of the model by EW physics. Red: $2\sigma$ (light) $3\sigma$ (dark) exclusion from $Zb_Lb_L$ vertex experimental input $\delta g_{Zb_Lb_L}/\delta g^{SM}_{Zb_Lb_L}=0.86\pm0.21$ \cite{lep2final}. Black: $10^3~\hat T$ contours. As an indication $-1.5\leq 10^3~ \hat T \leq 3.0$ at $2\sigma$, neglecting correlation with $\hat S$. Blue dashed: fine tuning amount quantified as in eq. \ref{finetuning}.
\label{EWPT}}
\end{figure}

In Fig. \ref{EWPT} we show the constraints on the parameter space of the model in the plane $(f,\lambda_1/\lambda_2)$ accounting for the various effects discussed up to now. We consider two values of the Higgs boson mass $m_h=115\GeV, 250\GeV$. In the same plane we give an estimate of the fine-tuning according to the formula
\be\label{finetuning}
F^{-1}\equiv\left|\frac{\delta m_h^2}{m_h^2}\right|\approx\frac{3 \lambda_t^2 m_T^2}{4\pi^2 m_h^2}\log\frac{\Lambda^2}{m_T^2}.
\ee
This is the log-divergent contribution of the top sector to the Higgs boson mass extracted from the CW potential \ref{CWlog}. Eq. \ref{finetuning} quantifies the statement that the particles which cutoff the quadratic divergences exhibited by SM have to be light in order for the theory to be natural. 

As anticipated, the model is consistent with EWPT in a large portion of its parameter space. This simplified analysis is in accord with the one in \cite{perel} where a global fit of the complete set of EW observable is performed. We stress that such comparison can be made since the lagrangian used in \cite{perel} coincide with the one of our model for the calculation of the relevant effects included in the fit.

Another constraint on the model come as an upper limit on $\kappa$ and so on the masses of the heavy $T$-odd fermions as a result of the upper bound on four-fermion operators discussed in section 2. A safe limit is \cite{perel}
\be
m_{Q^-}\leq 1.2 ~\tev \left(\frac{f}{500\GeV}\right).
\ee
Since EWSB doesn't result in a mass splitting for these heavy fermions, there's no extra contribution to $\hat T$ coming from them. \newline

One of the virtues of $T$-parity is that it provides a natural candidate to explain the cosmological DM abundance: the lightest $T$-odd particle (LTP). This particle is stable if $T$-parity is an exact symmetry\footnote{We leave for the concluding sesction a brief account of the observation that $T$-parity can be broken by anomalies\cite{hill2}.}. In \cite{littlestpheno, littlestDM} the LTP is identified with the heavy partner $B_H$ of the $B$ vector boson. Its mass is univocally fixed by the model to be $g' f/\sqrt 5$ up to $v/f$ corrections. The dominant annihilation mode is through $s$-channel Higgs boson exchange. In the region where the heavy doublet fermions are almost degenerate with the $B_H$ co-annihilation with them are also important.

In the model proposed in this paper the $B$ partner is still light enough to be a valid DM candidate. On the other hand there is a richer scalar sector that can provide an alternative realization of the framework proposed in \cite{scalarDM}. The neutral component of the triplet is not a good candidate if its mass lays in the natural range $\mathcal O(100\,\gev)$ dictated by the present model or if its mixing with the singlet is not substantial. If the triplet is above the $W^+W^-/ZZ$ threshold, which is a quite necessary requirement due to a conservative bound $m_{\varphi^\pm}>100\,\gev$ from LEP, then its mass is required to be in the $2-3\,\tev$ range \cite{mdm} to explain the observed DM abundance.

 A possibility for a viable DM candidate is thus the singlet scalar $s$. It's phenomenology is similar to that of the ``darkon'' already discussed in \cite{darkon}. At low masses the main annihilation mode is through $s$-channel higgs-boson exchange via the $shh$ coupling generated from the $s^2H^\dagger H$ term after EWSB. Subleading contributions come from $t$-channel heavy doublet fermion exchange, which are suppressed by the mass of the final state fermions. Our singlet scalar departs from the ``darkon'' at higher masses when the $sshh$ derivative interactions from the $\sigma$-model sets in. These act as an affective quartic coupling scaling as $m_s^2/f^2$ times a small numerical factor coming from the normalization of the generator associated to the singlet. We find that the $\sigma$-model contribution interferes destructively with the pure quartic coupling.
 
\begin{figure}[t] 
{\includegraphics[scale=0.8]{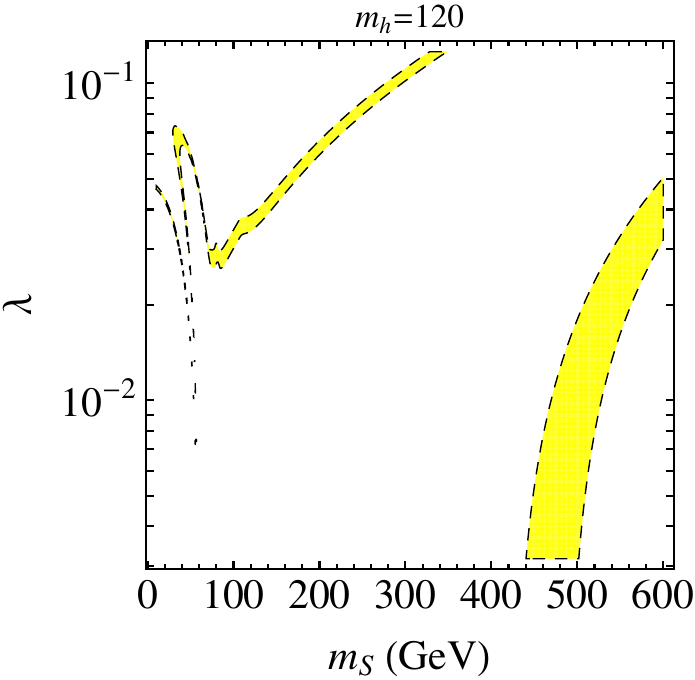}\includegraphics[scale=0.8]{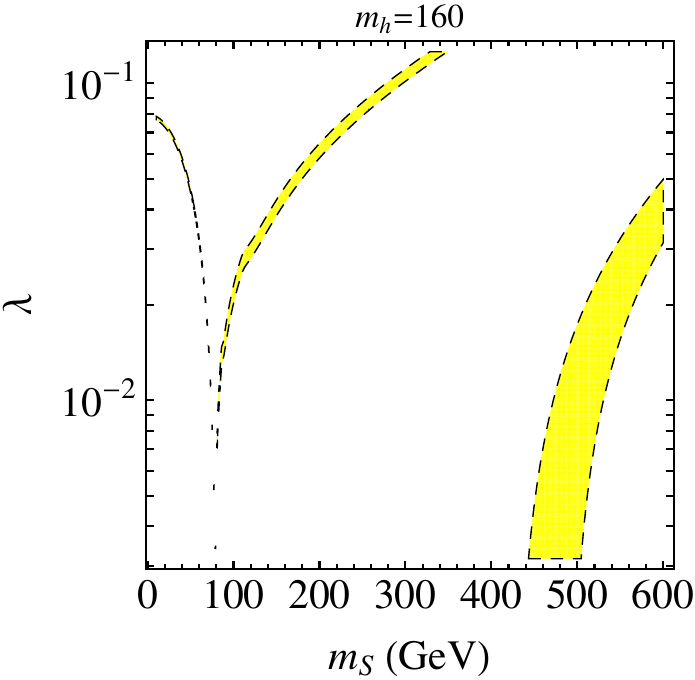}\includegraphics[scale=0.8]{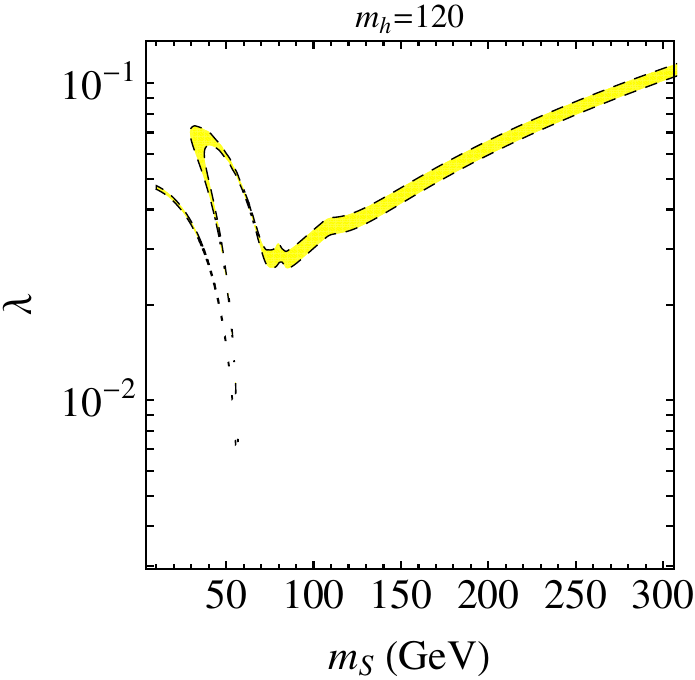}}
\caption{\small 3$\sigma$ allowed region for the relic abundance. In the plot the $\sigma$-model breaking scale $f$ is fixed at $1\,\tev$. To the experimental uncertainty, we add in quadrature a 5\% theoretical error to take care of neglected $p$-wave effects.} 
\label{DM}
\end{figure}

In fig.\ref{DM} we plot the region favored by the WMAP measurement \cite{WMAP}, $\Omega_{DM}h^2=0.110\pm0.005$, as a function of $m_S$, the mass of the singlet prior EWSB, and $\lambda$, the quartic $ssH^\dagger H$, for various values of the Higgs boson mass. In the low mass region the abundance is fixed by the coupling $\lambda$. For the values one naturally obtains in the model, $\lambda\sim$ few percent, the mass of the singlet is required to be close to $m_h/2$ where resonant Higgs boson production takes place\footnote{It is interesting to notice that if $m_s\lesssim m_h/2$ the Higgs width become easily dominated by the invisible mode $h\rightarrow ss$ \cite{strumia}}. Augmenting the mass for a fixed $\lambda$, brings into a region where the destructive interference among the quartic and the non-linear interactions makes the annihilation cross section too small.  This happens up to a value for $m_S$ where the correct relic abundance can be again reproduced. 
Such values for the singlet mass are probably too big w.r.t. their natural size expected from \ref{scalars}.


Various corrections have been neglected in fig.\ref{DM}. One of them is the mixing among the singlet and the triplet. A moderate mixing will generally require lower values of $\lambda$ to fit the correct DM abundance. A similar effect would be induced by the co-annihilations among the various scalars (singlet, triplet and also the Higgs boson) and possibly the $B_H$, if these particles are almost degenerate with $s$. These effects deserve a more careful evaluation.\newline

We conclude this section with a brief account of the collider phenomenology of the model. There is a vast literature \cite{pheno} studying the possible collider signatures of little Higgs models with T-parity. Pair production of heavy quark doublets $Q_iQ_j$ is promising, since these doublets are predicted to be many, one for each SM quark doublet, and light, to evade the LEP bounds on four-fermion operators mentioned in section 2. Summing on the flavors of the first two generations one can achieve cross sections around few picobarns  at LHC@14\tev ~for masses of the quarks up to $1\,\tev$ \cite{belyaev}. Roughly one order of magnitude lower is the cross section for the associated production $Q_i V_H$ of these heavy doublets together with the heavy vector bosons.

In the top sector, the heavy $T$-odd top partner is predicted to be lighter than its $T$-even relative according to \ref{toppartners} and will thus have a bigger pair production cross section (around 1pb at LHC@14\tev~for a 600\gev ~fermion). The $T$-even top partner can be singly produced in association with a SM vector boson. The cross section for this process turns out to be roughly equal to $T^-T^-$ pair production for a given choice of the model parameters.

Somewhat below these numbers one finds the pair production of heavy vector bosons and much below the pair production of the heavy triplet. 

What of course is relevant for the collider phenomenology is the convolution of these numbers with the relevant branching ratios and a careful evaluation of the SM backgrounds.
Even though a more precise study may be in order we believe that no major surprise will come from the modification we proposed. The addition of light scalars could in principle modify the branching ratios. 

$SU(2)$ gauge invariance and $T$-parity may help understanding the size of these corrections. In the top sector, $(t,T^+,T^-)$, at leading order in $v/f$, only a new coupling can appear: $\bar T^-t s$. Furthermore this coupling is weighted by the small normalization factor of the singlet generator.

More relevant modifications are expected for the branching ratios of the heavy $T$-odd doublets. These are tipically assumed to decay as $Q_i\rightarrow q_j V_H$ \cite{belyaev}. New couplings like $\bar Q_i q_j s$ or $\bar Q_i\sigma^I q_j\varphi^I$ are present in the model with strength fixed by $\kappa$ as in \ref{eq:oddcombination}. These interactions are expected to modify $\mathcal O(1)$ the usually assumed branching ratios giving rise to interesting signatures like chain decays among the new light scalar degrees of freedom, resembling the one found in supersymmetric models.


\section{Summary and conclusions}\label{conclusion}
Little-Higgs models are an ambitious attempt to solve the so called ``little hierarchy problem''. Taken at face value specific LH models encounter various difficulties in reaching this goal. $T$-parity is introduced to overcome these problems. Explicit model building shows that the implementation of $T$-parity is not straightforward in the fermionic sector. We gave an account of this in section \ref{tparityproblems}. We repeat here the logic.

The elementary nature of the SM fermions, accurately tested by LEP, forces to embed them in the theory as linearly transforming representations of the broken symmetry group $\mathcal G$. To satisfy $T$-parity invariance it is necessary to introduce at least a pair of electroweak doublets for each SM one. This in turn requires at least another doublet to supply a mass term to the extra fields. The only way out of this is the introduction of composite representations of the unbroken group $\mathcal H$, since they can be $T$-parity eigenstates. These representations, however, never contain just a single electroweak doublet. This is usually accompanied by other fields which in specific examples are extra singlets or doublets.  One has to deal with these extra states: chiral doublets cannot be massless, and electroweak singlets modify $Z$-pole observables. We remarked that composite incomplete representations are not a viable alternative and we also pointed out that other strategies that are commonly adopted are not working.

In this paper we proposed a strategy out of the above difficulties and we presented a low energy model that realizes it explicitly. 
We suggest to enlarge the symmetry breaking pattern and in particularly the unbroken symmetry group $\mathcal H$ in such a way to allow the introduction of a complete composite representation containing just one fermion doublet. The task of the new Goldstone bosons in the enlarged coset is to lift this representation to an elementary one and to connect it to the SM fermions in a gauge invariant manner (see eq.\ref{eq:oddcombination}). The specific way in which this is realized depends on the coset. We chose to build up starting from the littlest Higgs since this is the simplest model with a single Higgs doublet.
We are nevertheless confident that the above construction can be adapted to other models such as those discussed in \cite{tparity1} and \cite{tparity2}.

Compared to the usual littlest Higgs with $T$-parity our model contains in addition a real triplet and a real singlet, both $T$-odd, with electroweak scale masses protected by the collective breaking mechanism. The main phenomenological consequences of their presence are described in section \ref{phenomenology} and have to do with the possible role of the singlet as the dark matter.\newline

Finally a comment about the validity of $T$-parity as an exact symmetry is in order. Some doubts have been raised in the literature concerning the possibility for $T$-parity to be broken by anomalies \cite{hill2}, in analogy with the anomalous breaking of ``pion parity'' in the QCD chiral lagrangian by the Wess-Zumino-Witten (WZW) term \cite{wzw}. Such a breaking, if present,  is not worrisome for EWPT, but makes the LTP unstable and undo the natural explanation of the cosmological DM relic abundance provided by LHT models.

The existence of a non vanishing WZW term depends on the ultraviolet completion of the theory. If this is QCD-like it is foreseeable for some of the chiral symmetries to be anomalous \cite{yavin}. On the other hand this is obviously not a problem in a linear completion of which \cite{csaki} is a concrete example. A proposal has also been put forward \cite{yavin, wyler} to modify the implementation of $T$-parity in moose models in such a way for this new symmetry, dubbed $X$-parity, to be untouched by a non vanishing WZW term. 


\section{Acknowledgments}
We are grateful to Riccardo Rattazzi for his guidance. We thank Ian Low for discussion and comments on the manuscript. We thank Alessandro Strumia for hastening the delivery of the paper. This work is supported by the Swiss National Science Foundation under contract No. 200021-116372.

\appendix
\section{Spinorial representations and the littlest Higgs coset}
We want to justify here the assertion we made in section 2 about the impossibility to embed spinorial representations in a model which is based on the littlest Higgs coset. The following argument is given in the paper of Callan et al. \cite{CCWZ}.

A symmetry breaking pattern $\mathcal G/\mathcal H$ is fixed. Given a composite field $\psi$ transforming as a certain representation $\mathcal D$ of $\mathcal H$
\be
\psi'=\mathcal D(h)\psi, \qquad h\in \mathcal H
\ee
one wants to build a function $f$ of the Goldstone field $\xi$ in such a way that $f(\xi)\psi$ transforms linearly under $\mathcal G$ in a representation $\mathcal D^0$
\be\label{linearrep}
f(\xi')\psi'=\mathcal D^0(g)f(\xi)\psi,\qquad g\in \mathcal G.
\ee

In our specific case $\mathcal G$ is $SU(5)$, $\mathcal H$ is $SO(5)$ and $\mathcal D$ is the spinorial representation. We want to see whether it is possible to lift $\mathcal D$ to a linear representation $\mathcal D_0$ of $SU(5)$.

A constraint on the linear representations that can be obtained this way is that once $\mathcal D_0$ is reduced with respect to $\mathcal H$, it must contain $\mathcal D$ as a sub-representation. The proof is  an application of the Schur's lemma. Taking $g=h\in \mathcal H$ and $\xi=0$ in \ref{linearrep}, $\xi'=0$ and
\be
f(0)\mathcal D(h)\psi=\mathcal D^0(h)f(0)\psi.
\ee
By Schur's lemma either $f(0)=0$ or $\mathcal D^0$ contains $\mathcal D$ as a sub-representation. The first possibility is excluded since would imply $f=0$ identically.

This shows that once the coset of the littlest Higgs is fixed no composite spinorial representation can be added consistently with the symmetry structure. Indeed given the $SO(5)\subset SU(5)$ embedding of the littlest Higgs, no $SU(5)$ representation decomposes in such a way to contain a $\mathbf 4$ of $SO(5)$.


\section{The complete Lagrangian}

The complete Lagrangian of the model is the sum of four pieces:
\ba
\	&&\mathcal L=\mathcal L^{\text{gauge}}+\mathcal L^{\text{nl$\sigma$m}}+\mathcal L^{\text{ferm}}+\mathcal L^{\text{pot}}\,,\\
\	&&\mathcal L^{\text{gauge}}=\sum_{i=1,2}-\frac{1}{4}W^a_{i\mu\nu}W^{a\,\mu\nu}_i-\frac{1}{4}B_{i\mu\nu}B^{\mu\nu}_i-\frac{1}{4}G^a_{\mu\nu}G^{a\,\mu\nu}\,,\\
\	&&\mathcal  L^{\text{nl$\sigma$m}}=\frac{f^2}8\text{Tr}[D\Sigma^* D\Sigma]+ \frac{f^2}{4}\text{Tr}[D{X^2}^\dagger DX^2]\,,
\ea
where the covariant derivatives for the linearly transforming fields are defined as
\ba
D\Sigma&=&\partial \Sigma - i \sum_{i=1,2} g_i W^a_i\left(Q_i^a \Sigma + \Sigma Q_i^{aT}\right)- i \sum_{i=1,2} g'_i B_i\left(Y_i \Sigma + \Sigma Y_i^{T}\right)\,,\\
D X^2&=&\partial X -\frac{i}2 \left(g_1 W^a_1 \sigma^a X^2-g_2  W^a_2 X^2\sigma^a\right)+\frac{i}{10}g'_1 B_1  X^2-\frac{i}{10}g_2'B_2 X^2 \,,
\ea
and $g_1=g_2=g\equiv \sqrt 2\,g_{SM}$ and $g_1'=g_2'=g'\equiv \sqrt 2\,g_{SM}'$ due to T-parity. The embedding of the generators $Q_i^a$ inside $SU(5)$ is defined in eq. \ref{generators} and $\sigma^a$ are the Pauli matrices. The choice of the charges appearing in the covariant derivative of $X^2$ will be justified later.\\
The parametrization of the Goldstone bosons inside the matrices $\Sigma$ and $X$ is reported in eq. \ref{goldstoni}. 
We recall the transformation properties of the Goldstone matrices $\xi=e^{i\Pi_\Sigma/f}$ and $X=e^{i\Pi_X/f}$ under the global symmetries:
\ba\label{eq:Goldstonetransformations}
\	SU(5):&&\xi\longrightarrow g \,\xi\, h^\dagger=h\, \xi \,\widetilde g^\dagger\\
\	SU(2)_{L}\times SU(2)_R:&&X\longrightarrow U_L \, X\, V^{\dagger}= V \, X\, U_R^\dagger\\
\	U(1)_{L}\times U(1)_R:&&X\longrightarrow e^{\frac{i}{2}(\alpha_L-\alpha_R)} \, X=e^{i\alpha_L}X e^{-\frac{i}{2}(\alpha_L+\alpha_R)}= e^{\frac{i}{2}(\alpha_L+\alpha_R)}Xe^{-i\alpha_R}\,\,\, 
\ea
where $g\in SU(5)$, $h$ is a non linear transformation $\in SO(5)$, $U_{L(R)}\in SU(2)_{L(R)}$ and $V$ is a non  linear transformations $\in SU(2)_{L+R}$. Finally $\alpha_{L(R)}$ are the parameters associated to $U(1)_{L(R)}$ transformations. Notice that, because of the abelian nature of the broken $U(1)_{L-R}$, the analogous of the non linear transformation $V$ is played by $e^{\frac{i}{2}(\alpha_L+\alpha_R)}$.

For the fermionic lagrangian we distinguish the leptons and the first two generations of quarks from the third generation. For each lepton $\ell$ we introduce a $\mathbf 5^*$, a $\mathbf 5$ both incomplete of $SU(5)$ and left-handed, a composite right-handed doublet of $SU(2)_V$ and a right-handed singlet:
\ba
\	&&\Psi^\ell_{1}=\begin{pmatrix}\psi^\ell_{1}\\0\\0\end{pmatrix}\,,\quad
	\Psi_{2}^\ell=\begin{pmatrix}0\\0\\ \psi^{\ell}_{2}\end{pmatrix}\,,\quad\Psi_{C}^\ell\,,\qquad
	\ell_R\,,\quad \ell=e,\mu,\tau\,.
\ea
The same holds  for the first two quark generations with the addition of extra right-handed singlets to give mass to the up-type quarks: 
\ba
\	&&\Psi_{1}^q=\begin{pmatrix}\psi_{1}^q\\0\\0\end{pmatrix}\,,\quad
	\Psi_{2}^q=\begin{pmatrix}0\\0\\ \psi_{1}^q\end{pmatrix}\,\quad\Psi_{C}^q\,,\quad
	u^q_R\,,\quad
	d^q_R\,,\quad q=1,2\,.
\ea
For the top quark more fields are needed as explained in Section 3:
\be
\mathcal T_1 =\begin{pmatrix}\psi^t_{1}\\ \chi_1 \\0\end{pmatrix},\qquad \mathcal T_{2} =\begin{pmatrix}0\\ \chi_2 \\ \psi^t_2\end{pmatrix},\qquad \mathcal T_C,  \qquad \tau_{1,2},\qquad t_R\,,\quad b_R\,.
\ee
The complete fermionic lagrangian reads
\ba
\	&& \mathcal L^{\text{ferm}}= \mathcal L^{\text{Kin}}+\mathcal L^{\text{mass}}+ \mathcal L^{\text{Yuk up}}+ \mathcal L^{\text{Yuk down}}+ \mathcal L^{\text{Yuk top}}\,,
\ea
where
\ba
\	&&\mathcal L^{\text{Kin}}=\sum_{j=1,2}\left(\sum_{f=\ell,q}i\overline\Psi^f_{j}\displaystyle{\not}D \Psi^f_j+i\overline{\mathcal T}_{j}\displaystyle{\not}D \mathcal T_j+i\overline{\tau}_j\displaystyle{\not}D\tau_j\right)+\sum_{f=\ell,q}i\overline\Psi^f_{C}\displaystyle{\not}D \Psi^f_C\nonumber\\
\	&&\phantom{......}+\sum_\ell i\overline{\ell}_R\displaystyle{\not}D \ell_R+\sum_qi\overline{u}_R^q\displaystyle{\not}D u_R^q+i\overline{d}_R^q\displaystyle{\not}D d_R^q+i\overline{t}_R\displaystyle{\not}D t_R+i\overline{b}_R\displaystyle{\not}D b_R\,.
\ea
Fermions are charged under an additional unbroken $U(1)_f$ which is needed to adjust the hypercharges:
\ba
\	&& D_\mu\Psi_1^f=\left(\partial_\mu +i g W_{1\mu}^aQ_1^{a\,T} +ig W_{2\mu}^aQ_2^{a\,T}+ig' B_1Y_1 +i g'B_2Y_2-i g'q_f (B_1+B_2)\right)\Psi_1^f\,,\qquad\qquad\\
\	&& D_\mu\Psi_2^f=\left(\partial_\mu -i g W_{1\mu}^aQ_1^{a} -ig W_{2\mu}^aQ_2^{a}-ig' B_1Y_1 -i g'B_2Y_2-i g'q_f (B_1+B_2)\right)\Psi_2^f\,.
\ea
The above covariant derivatives apply also to the $\mathcal T_j$ fields. We can collectively denote $q_1$  and $q_2$ the couplings to the $U(1)$ gauge bosons $B_1$ and $B_2$; for fermions embedded into some representation of $SU(5)$ we have the relation:
\be
	q_1=Y_1+q_f\,,\qquad q_2=Y_2+q_f\,,\qquad Y=q_1+q_2\,.
\ee
where $Y_{1,2}$ are the two diagonal generators in eq. \ref{generators}.
The covariant derivatives for the singlets read $\partial_\mu -i g' q_1 B_1-i g' q_2 B_2$. Quarks and leptons have different values of $q_1$ and $q_2$ in order to reproduce the correct SM hypercharge.
All the $U(1)$ charges are collected in Table. \ref{hypercharges}.
\begin{table}[h]
\centering
\begin{tabular}{|c|c|c|c|c|c|c|}
\hline
 &$Y_1$&$Y_2$&$q_f$&$q_1$&$q_2$&$Y$	\\
 \hline\hline
 $\psi_1^q $&$-3/10 $&$-1/5$&$1/3$&$1/30$&$2/15$&$1/6$\\
 $\psi_2^q $&$-1/5 $&$-3/10$&$1/3$&$2/15$&$1/30$&$1/6$\\
 $u^q_R $&$ 0$&$0$&$1/3$&$1/3$&$1/3$&$2/3$\\
  $d^q_R $&$ 0$&$0$&$-1/6$&$-1/6$&$-1/6$&$-1/3$\\
   $\chi_1 $&$1/5 $&$-1/5$&$1/3$&$8/15$&$2/15$&$2/3$\\
   $\chi_2 $&$-1/5 $&$1/5$&$1/3$&$2/15$&$8/15$&$2/3$\\
   $\tau_1 $&$0 $&$0$&$$&$8/15$&$2/15$&$ 2/3 $\\
   $\tau_2 $&$0 $&$0$&$$&$2/15$&$8/15$&$ 2/3 $\\
 $\psi_1^\ell $&$-3/10 $&$-1/5$&$1/2$&$1/5$&$3/10$&$-1/2$\\
$\psi_2^\ell $&$-1/5 $&$-3/10$&$1/2$&$3/10$&$1/5$&$-1/2$\\
  $\ell_R $&$0 $&$0$&$-1/2$&$-1/2$&$-1/2$&$-1$\\
  \hline
\end{tabular}
\caption{\small $U(1)$ charge assignments to fermion fields. Since $\tau_{1,2}$ have different charges under $B_1$ and $B_2$, it is not possible to define $q_f$.}
\label{hypercharges}
\end{table}

To conclude we write the covariant derivatives for $\mathcal T_C$ and $\Psi_C^f$. According to the CCWZ  construction \cite{CCWZ} they must contain the Goldstone bosons and, given the transformation laws (\ref{eq:Goldstonetransformations}), we have
\ba
\	&& D_\mu T_C=\partial_\mu T_C+\frac{1}2 \left(X^\dagger D_\mu X +XD_\mu X^\dagger\right)\mathcal T_C -i g'q_f (B_1+B_2)\mathcal T_C\,,\\
\	&& D_\mu X\equiv \partial_\mu X -i g W_{1\mu}^a \tau^a X +\frac{ig'}{10} B_1X\,,\\
\	&& D_\mu X^\dagger \equiv \partial_\mu X^\dagger -i g W_{2\mu}^a \tau^a X^\dagger +\frac{ig'}{10} B_2 X^\dagger\,.
\ea
where we have allowed for a $U(1)_f$ charge. In order to determine the latter we can impose the gauge invariance of the interaction terms $\bar{ \mathcal T}_1\sigma_2 X \mathcal T_C+ \bar{ \mathcal T}_2\sigma_2X^\dagger \mathcal T_C$. Considering for instance a transformation associated to the gauge field $B_1$ we get the conditions:
\be
\	-\frac{1}{30}-\frac{1}{10}+q_f=0 \,\qquad -\frac{2}{15}+q_f=0
\ee
The above equations are satisfied (together with those coming from the gauge transformation associated to $B_2$) for the choice $q_f=2/15$. In the case of leptons the choice would be $q_f=-1/5$.\\
We emphasize the importance of the factor $-1/10$ in the above equation: it shows the necessity of having a broken $U(1)$ factor under which the field $X$ transforms. If this were not the case the above equations would not allow for a solution.

The remaining pieces of the fermionic Lagrangian contain the terms needed to provide a large mass to the $T$-odd combination of the doublets and the standard Yukawa interactions for leptons, light quarks and the top quark:
\be
\mathcal L^{\text{mass}}=-\sum_{f=\ell,q} \frac{\kappa f}{\sqrt 2}\left(\overline\psi_{1}^f\sigma_2 X+\overline\psi_{2}^f\sigma_2 X^\dagger \right)\Psi_{C}^f\,
\ee
assuming an universal $\kappa$.
The Yukawa terms for the light fermions can be introduced without taking care to preserve larger global symmetries since the only corrections to the Higgs boson mass that are generated are proportional to the small fermion masses. We have then:
\ba
\	\mathcal L^{\text{Yuk up}}=\sum_{q=1,2} \frac{\lambda_f}{4}\epsilon_{ab}\epsilon_{ijk}\left(\overline\Psi_{1\,i}^q \Sigma_{ja}\Sigma_{kb}-(\overline\Psi_{2}^q\Sigma_0\Omega)_{i} \widetilde\Sigma_{ja} \widetilde\Sigma_{kb}\right)u^q_R\,, \qquad  \widetilde\Sigma\equiv \Omega\Sigma_0 \Sigma^\dagger\Sigma_0\Omega\nonumber\\
\ea
The above term reduces to the usual Yukawa interaction once expanded around the electroweak VEV
\be
\	\mathcal L^{\text{Yuk up}}\rightarrow \sum_{q=1,2}-i\frac{\lambda_q}{\sqrt2}(\bar\psi_1^q-\bar\psi^q_2)H^*u^q_R=\sum_{f=q}-\lambda_q\epsilon_{ab}\bar q_{a} \,H^*_b \,u^q_R
\ee
where we made use of the identification:
\ba
\	\frac{\psi_1^q-\psi_2^p}{\sqrt2}\equiv \sigma_2 q\equiv\sigma_2 \begin{pmatrix}u_L^q\\ d_L^q\end{pmatrix}\,.
\ea
The Yukawa interactions for the down fermions are more involved due to the required invariance under the $U(1)$'s:  
\ba
\	&& \mathcal L^{\text{Yuk down}}=\sum_{f=l,q} i\frac{\lambda'_f}{2}\epsilon_{ab}\epsilon_{ijk}\left(\Sigma_{33}^{-1/4}(\bar\Psi_2^f Q_V^2)_{i}) \Sigma_{aj}\Sigma_{bk}-\widetilde\Sigma_{33}^{-1/4}(\bar\Psi^f_{1}\Sigma_0 Q_V^2)_i \widetilde\Sigma_{ja} \widetilde\Sigma_{bk}\right)f_R\,, \nonumber\\
\	&& Q_V^a\equiv Q_L^a+Q_R^a\,.
\ea
In the above expression the presence of $\Sigma_{33}^{-1/4}$ is needed to achieve gauge invariance. Note that the above sum contains also the $b$-quark Yukawa interaction. Finally we report the Yukawa interaction for the top quark:
\be
\mathcal L^{\text{Yuk top}}=\frac{\lambda_1 f}{4}\epsilon_{ij}\epsilon_{abc}\left(\Sigma_{ai}\Sigma_{bj}\overline{\mathcal T}_{1c}  - \widetilde\Sigma_{ai} \widetilde\Sigma_{bj}(\overline{\mathcal  T}_2\Sigma_0\Omega)_c \right)t+\frac{\lambda_2 f}{\sqrt 2} (\overline \chi_1\tau_1+\overline \chi_2\tau_2).
\ee

\end{document}